\documentclass[handcarry,10pt]{nasa}
\usepackage{amssymb}
\usepackage{lineno}
\usepackage{graphicx}
\usepackage{tabularx}
\usepackage{url}
\usepackage[american]{babel}
\usepackage{amsmath,amsfonts}
\usepackage{algorithmic}
\usepackage{graphicx}
\usepackage{textcomp}
\usepackage{xcolor}
\usepackage{mathtools}
\usepackage{booktabs} 
\usepackage{tikz} 
\usepackage{relsize}
\usepackage{subcaption}
\usepackage{booktabs}
\usepackage{siunitx}
\usepackage[makeroom]{cancel}
\usepackage{MnSymbol}
\usepackage{algorithm}
\usepackage{float}
\usepackage{dblfloatfix}
\usepackage[section]{placeins}
\usepackage{enumitem}
\usepackage{hyperref}
\usepackage{xcolor}

\makeatother
\title{Robust Design in the Presence of Aleatoric and Epistemic Uncertainty}
\author {Luis G. Crespo\thanks{MS 308, NASA LaRC, Hampton VA 23681 USA.} \\
         {\normalsize\itshape Vehicle Dynamics and Control Branch, NASA Langley Research Center}
       }

 \newcommand{\argmin}{\operatornamewithlimits{argmin}}

\newcommand{\Pro}{{I\kern-.35em P}}

\DeclarePairedDelimiter\floor{\lfloor}{\rfloor}

\newtheorem{theorem}{Theorem}

\newtheorem{definition}{Definition}

\begin{document}
\maketitle
\begin{abstract}\noindent This paper proposes strategies for designing a system whose computational model is subject to aleatory and epistemic uncertainty. Aleatory variables, which are caused by randomness in physical parameters, are draws from a possibly unknown distribution; whereas epistemic variables, which are caused by ignorance in the value of fixed parameters, are free to take any value in a bounded set. Chance-constrained formulations enforcing the system requirements at a finite number of realizations of the uncertain parameters are proposed. These formulations trade off a lower objective value against a reduced robustness by eliminating an optimally chosen subset of such realizations. Risk-aware designs are obtained by accounting for the severity of the requirement violations resulting from this elimination process. Furthermore, we propose a computationally efficient design approach in which the training dataset is sequentially updated according to the results of high-fidelity reliability analyses of suboptimal designs. Robustness is evaluated by using Monte Carlo analysis and Robust Scenario Theory, with the latter approach accounting for the infinitely many values that the epistemic variables can take. 
\end{abstract}

\section{Introduction} 
Optimization under uncertainty addresses the challenge of making optimal decisions about a system whose performance depends on parameters whose value might either vary or be unknown. Uncertainty might be caused by randomness in the physical system, ignorance in the computational model, or by changing operating conditions. Traditional deterministic optimization assumes that all the parameters of such a model are known, a rare case in real-world applications. Optimization under uncertainty, on the other hand, is a decision-making process that accounts for the effects of uncertainty upfront.  A plethora of applications in finance, science and engineering \---including  regression analysis, model calibration, controls, system identification, machine learning, systems theory, financial mathematics, structural design\---  require decision-making under uncertainty.

The mathematical framework underpinning the proposed strategies is introduced next. Let $J:\Theta \rightarrow \mathbb{R}$ be an objective function depending on the decision variable $\theta\in{\mathbb R}^{m_\theta}$. Furthermore, let $r_k(\theta,a,e): \mathbb{R}^{m_{\theta}} \times \mathbb{R}^{m_{a}} \times \mathbb{R}^{m_e} \rightarrow  \mathbb{R}$ for $k\in1,\ldots n_r$ be a continuous function associated with a system requirement, where  $a\in \mathbb{R}^{m_a}$ is a parameter subject to {\em aleatory} uncertainty, and $e\in \mathbb{R}^{m_e}$ is a parameter subject to {\em epistemic} uncertainty. 
It will be assumed that the $k$-th requirement is satisfied when $r_k(\theta,a,e)\leq 0$. Whereas the value $a$ varies randomly according to the possibly unknown distribution $\mathbb{P}_a$ supported in $A \subset \mathbb{R}^{m_a}$, the true value of $e$ is an unknown constant taking any value in $E \subset \mathbb{R}^{m_e}$. In contrast to the aleatoric uncertainty, experiments and/or simulations can be used to reduce epistemic uncertainty, i.e., to reduce the size of $E$. 

Consider the optimization program 
\begin{align}
	 \min_{\theta \in \Theta} & \quad J\left(\theta  \right) \label{sta}\\
	\text{ subject to:} & \quad \mathbb{P}_a[r_k(\theta,a,e)> 0]\leq \alpha_{k} \;\;\text{for all}\;\; e\in E,\;  k=1,\ldots n_r, \nonumber 
\end{align}
where 
$0\leq \alpha_k \ll 1$ is the maximally acceptable probability of violating the $k$-th requirement. Denote as $\theta^\star$ the solution to (\ref{sta}). Hence, we seek a decision point (or design) that minimizes the objective function $J(\theta)$ such that the greatest probability of violating the $k$-th requirement is acceptably small. This optimization program, which has $n_r$ semi-infinite \emph{Chance Constraints} (CC), is generally intractable. 

The \emph{feasible set} of (\ref{sta}) corresponding to the aleatory point $a\in A$ is 
\begin{equation}
\label{feaset}
\mathcal{X}(a, E)=\bigcap_{e\in E} \left \{\theta\in \Theta:  r_{\text{max}}(\theta,a,e)\leq 0 \right\},
\end{equation}
where
\begin{equation}
r_{\text{max}}(\theta,a,e):= \max_{k=1,\ldots n_r} r_k(\theta,a,e)\label{wcr}
\end{equation}
is the Worst-Case (WC) requirement function. Hence, $\mathcal{X}(a, E)$ is comprised of the decision points that satisfy the WC requirement for the fixed aleatory point $a$ and all the epistemic points\footnote{Program (\ref{sta}) is a conservative approximation to the joint CCed program
\begin{align}
	 \min_{\theta \in \Theta} & \quad J\left(\theta  \right) \label{staj}\\
	\text{ subject to:} & \quad \max_{e\in E}\; \mathbb{P}_a\left [ r_{\text{max}}(\theta,a,e)>0\right ]\leq \alpha, \nonumber 
\end{align}
when $\alpha > \sum_{k=1}^{n_r} \alpha_k$. Conservative approximations have a smaller feasible set, so they might lead to a higher $J(\theta^\star)$.  Requirement functions exhibiting strong parameter dependencies and suboptimal choices for the $\alpha_k$'s increase the conservatism. 
A joint CC can be cast in terms of individual CCs without introducing any conservatism.} in $E$.

The \emph{success domain} of (\ref{sta}) corresponding to the design $\theta \in \Theta$ is
\begin{equation}
\mathcal{S}(\theta,E)=\bigcap_{e\in E}\left \{a\in A:\; r_{\text{max}}(\theta,a, e)\leq 0 \right\}.\label{su}
\end{equation}
Hence,  $\mathcal{S}(\theta,E)$ is comprised of all the aleatory points for which $\theta$ satisfies the requirements for all the epistemic points in $E$. The aleatory space is partitioned into the success domain and the \emph{failure domain} $\mathcal{F}$, so $\mathcal{F}(\theta,E)\cup \mathcal{S}(\theta,E)=\mathbb{R}^{m_a}$ and $\mathcal{F}(\theta,E)\cap \mathcal{S}(\theta,E)=\emptyset$.  The $k$-th individual feasible set, $\mathcal{X}_k(a, E)$, individual success domain, $\mathcal{S}_k(\theta,E)$, and individual failure domain, $\mathcal{F}_k(\theta,E)$, result from replacing $r_{\text{max}}$ with $r_k$ in the above expressions. 
  
The field of optimization under uncertainty was pioneered in the 1950s by Dantzig \cite{Dantzig55} and Charnes \cite{Charnes59}, who set the foundations for \emph{Robust Optimization} (RO) and \emph{Probabilistic Optimization} (PO), respectively. RO seeks designs for which $\max_{e\in E} r_k(\theta,e)\leq 0$. Conversely, PO seeks designs given by (\ref{sta}) where $E$ is a singleton. In contrast to these formulations, (\ref{sta}) depends on aleatoric and epistemic uncertain parameters. 

The main advantage of RO designs is the robust feasibility guarantee:  the requirements are satisfied for all epistemic points in $E$. However, general robust problems are often intractable even when the requirement functions are convex, e.g., the robust counterpart of a second order cone program with polyhedral uncertainty is NP hard  \cite{Elghaoui98,Ben-Tal00,Bertsimas04}. Tractable robust linear optimization problems, which require a suitably chosen $E$ and particular requirement function forms, are summarized in \cite{Ben-Tal09, Bertsimas11}. And even when RO is tractable, the resulting design commonly underperforms other designs for most elements of $E$. Another drawback of RO is the high sensitivity of the resulting design to the set $E$ assumed upfront\footnote{The set $E$ should be prescribed according to the analyst's belief of where the true value of $e$ might be. While overly small sets might lead to designs that violate the requirements in practice, overly large sets might lead to under-performing conservative designs.}. 
Evaluating the extent by which epistemic uncertainty impacts the ability of a design to satisfy the system requirements will be called a \emph{robustness analysis}. 

PO prescribes the uncertainty as a distribution. This distribution is inferred from data, expert opinion or a mixture of these. Distributions enable the analyst to consider moments of a response function \cite{Tsatsanis98,Mehlawat21,Hammond24} as well as \emph{Chance-Constraints} (CC). PO programs subject to CCs are solved by \emph{Chance-Constrained Optimization} (CCO) \cite{Shapiro09,Coit19}. In contrast to RO, CCO seeks a design that allows for some requirement violations.  Most CC approaches either ignore epistemic uncertainties or treat them as if they were aleatory \cite{Shapiro09,Coit19}, thereby leading to designs that are either too conservative or insufficiently robust in practice. CC designs mitigate the intrinsic conservatism of RO designs by ignoring some of the aleatoric realizations in exchange for a lower objective value.  Calculating the failure probability at the core of a CC entails computing a high dimensional integral over a complex integration domain, a problem that is strongly NP-hard \cite{Luedtke10, Qiu14}.  Furthermore, the feasible set of CCO programs is often non-convex, making the identification of the global optimum difficult.  In fact, the feasible set of a CCed program corresponding to a convex constraint might be non-convex.\footnote{The convexity of the requirements is preserved by reformulating the CC in terms of the {\em Conditional Value at Risk} (CVaR) \cite{Shapiro09,Chapman22} at the expense of reducing the feasible set.} CCO programs rendering a convex feasible set can be efficiently solved \cite{Kataoka63, Lagoa99, Calafiore06, Henrion08, Henrion11, Prekopa11, Van15}. However, most practical engineering problems do not take such forms. A variant of CCO, called \emph{Reliability-Based Design Optimization} (RBDO), is commonly used in engineering. RBDO methods \cite{ZIO2009125, COIT2019106259} seek designs having an acceptably low failure probability for the fixed epistemic point $e_{\text{nom}}$. The standard approach to RBDO involves two nested loops: an outer loop that searches for an optimal design, and an inner loop that evaluates $\mathbb{P}_a[{\mathcal F}(\theta,e_{\text{nom}})]$ for every candidate design chosen by the outer loop \cite{ENEVOLDSEN1994169}. Evaluating the extent by which aleatory uncertainty impacts the ability of any given design to satisfy the system requirements will be called \emph{reliability analysis}. 

The high computational cost of accurately estimating small probabilities makes RBDO expensive \cite{8272792}. Decoupling approaches \cite{YUAN2014107}, single-loop methods \cite{SHAN20081218, MENG201995, WANG2020113436}, and efficient approximations to the failure probability have been used to make RBDO more efficient. Single-loop methods combine the outer and inner loops by substituting the reliability analysis with an approximation \cite{YAO201328}, whereas decoupling methods replace the nested optimization with a sequence of deterministic programs \cite{TORII2019106499, ChenSORA2002}. Methods that reduce the computational cost of the inner loop include subset simulation \cite{LI2010384}, line sampling \cite{DEANGELIS2015170}, importance sampling \cite{CHAUDHURI2020106853}, the first-order reliability method (FORM), the second-order reliability method (SORM), \cite{ENEVOLDSEN1994169, NIKOLAIDIS1988781, Kuschel1997, Cizelj1994, SCHUELLER2004463}, multi-fidelity surrogate-modeling strategies  \cite{LI20108966, CHEN2013233, PEHERSTORFER201761,doi:10.1137/17M1122992}  and many others  \cite{DIGE2018431, LI2019106432, Ullmann2015MultilevelEO}.  

A drawback of the CCO approaches is the high sensitivity of the resulting design to the distribution $\mathbb{P}_a$ assumed upfront. This prescription generally involves prescribing a distribution, a dependency structure, and a tail model from data. This process is challenging when the number of aleatory variables is large, when parameter dependencies are strong, when the dataset is incomplete, e.g., when the dependency between disjoint subsets of data cannot be measured, or when the data are scarce. Poorly chosen uncertainty models might lead to designs that grossly underperform in practice \cite{ROCKAFELLAR2010499, Sarykalin2008,LEBRUN2009172, LEBRUN2009312}. 
The subjectivity in prescribing a distribution can be mitigated by using distributionally robust approaches, which seek designs that satisfy the requirements for a family of distributions \cite{ELDRED20111092, Xie_DRCCP_Wassestein2018, NANNAPANENI20169}.  
 
The optimization program (\ref{sta}) combines aleatory and epistemic uncertainties, thereby making standard RO and CCO approaches inapplicable. The main difficulties in solving (\ref{sta}) stem from the inability to evaluate CCs with high accuracy and from having to account for the infinitely many elements in $E$.  The joint effect of both uncertainty types is often studied in a nested loop in which the outer loop accounts for the epistemic uncertainty whereas the inner loop accounting for the aleatory uncertainty. 
Sets, distributions, sets of distributions (possibility theory) and distributions of sets (Dempster-Shafer theory) have been used for modeling aleatory and epistemic uncertainties \cite{Zaman17,Mourelatos06,Youn08}. 
While aleatoric uncertainty is appropriately modeled by distributions, one may argue that a set-based approach is more suitable for modeling ignorance, and hence more apt at capturing epistemic uncertainty \cite{Hullermeier21}. Design in the presence of mixed uncertainty yields three nested loops: an outer loop for the decision variable, a middle loop for the epistemic uncertainty, and an inner loop for the aleatoric uncertainty. 

In this paper we seek data-driven approximations to (\ref{sta}).  Existing data-driven methods for CC optimization, which ignore epistemic uncertainties, include the scenario approach \cite{Calafiore06,Campi21,Paulson21,Campi21ml,Garatti22,Crespo25}, the sample average approximation \cite{Luedtke08}, and RO over safe approximations of the feasible set  \cite{Nemirovski07}.  Developments in the latter two categories, however, have mostly focused on convex requirement functions taking particular forms \cite{Geng19,Mammarella22}. The scenario strategies below, which do not suffer from such restrictions, are readily applicable to 
implicit requirement functions that must be evaluated by numerical simulation.

\section{Solution Strategy and Goals}
 The strategies below make direct use of a finite number of realizations of both $a$ and $e$. 
The realizations of the aleatory uncertainty, called \emph{aleatory scenarios}, comprise the training dataset\footnote{The scenarios in ${\mathcal A}$ should ideally be obtained experimentally through measurements. When experimental data is not available, they could be obtained synthetically through the simulation of an assumed distribution.} 
\begin{equation}
{\mathcal A}=\left \{a^{(i)}\right \}_{i=1}^{n_a},\label{ds}
\end{equation}
whereas the realizations of the epistemic uncertainty, called \emph{epistemic scenarios}, comprise the training dataset 
\begin{equation}
{\mathcal E}=\left \{e^{(j)}\right \}_{j=1}^{n_e}.\label{dse}
\end{equation}
The elements of ${\mathcal A}$ are observations drawn from a possibly unknown distribution $\mathbb{P}_a$, whereas those in ${\mathcal E}$ are drawn synthetically from the set $E'$. 
The scenarios in  $E'$ can be chosen at will. For instance, the analyst might choose scenarios uniformly distributed in $E'=E$ or use the conservative representation $E' \supset E$.  As expected, there is no guarantee that any training set having a finite number of scenarios will lead a design $\theta^\star$ satisfying the CCs in (\ref{sta}). However, a formal robustness analyses will be used to evaluate the probability of future data meeting the requirements for the infinitely many epistemic points $e$ in $E$. 

The goals of this article are threefold. First, we want to find an approximate solution to (\ref{sta}) by solving a scenario program that meets the requirements for a fraction of the scenarios in ${\mathcal A}$ and ${\mathcal E}$. By making this fraction as large as possible, we obtain the most robust design and a comparatively large $J(\theta^\star)$.  However, this value can be purposely lowered by considering a smaller fraction. In this fashion, performance and robustness can be systematically traded off. Second, we want to synthesize designs meeting the desired specifications while keeping the computational cost low. These goals are attained by sequentially expanding training datasets (\ref{ds}) and (\ref{dse}). Each iteration entails solving a scenario program having small training datasets (so the middle and inner loops are evaluated with low-fidelity), evaluating the resulting design using large testing sets, and then augmenting the training datasets with a small subset of carefully chosen testing scenarios. 
Finally, we want to evaluate the risk of the data-driven design $\theta^\star$ not meeting the desired specifications for future data. This determination is done statistically by using Monte Carlo analysis and formally by using robust Scenario Theory.  The reliability and robustness properties of the
designs resulting from this new framework are fundamentally different from those resulting from existing RO and PO formulations.  

\subsection{Risk-aware Design}
The scenario program at the core of the forthcoming formulations is
\begin{align}
	 \min_{\theta \in \Theta} & \quad J\left(\theta \right) \label{sta2}\\
	\text{ subject to:} & \quad r_k(\theta,a^{(i)}, e^{(j)})\leq 0,\; a^{(i)}\in \tilde{{\mathcal A}},\; e^{(j)}\in\tilde{{\mathcal E}},\;  k=1,\ldots n_r,\nonumber 
\end{align}
where the sets $\tilde{{\mathcal A}}\subseteq {\mathcal A}$ and $\tilde{{\mathcal E}}\subseteq {\mathcal E}$ are yet to be prescribed. The aleatoric scenarios ignored, ${\mathcal O}_a={\mathcal A} \setminus \tilde{{\mathcal A}}$, will be called the \emph{aleatoric outliers}, whereas the epistemic scenarios ignored,  ${\mathcal O}_e={\mathcal E} \setminus \tilde{{\mathcal E}}$, will be called the \emph{epistemic outliers}. The elimination of outliers mitigates the detrimental effects that extreme data points often have on data-driven designs \cite{Archimbaud18,Liang20,Guan19}. The scenarios in $\tilde{{\mathcal A}}$ and $\tilde{{\mathcal E}}$ will be called \emph{inliers}.

The process by which ${\mathcal O}_a$ and ${\mathcal O}_e$ are chosen might depend on the severity of the requirement violation. This feature is explained next. Risk is commonly defined as the probability of an adverse outcome times the corresponding ``loss'' or consequence. In the context of this paper, the adverse outcome is the violation of a requirement, whereas the loss measure is the corresponding positive value taken by $r_k$. Designs $\theta^\star$ that depend on a loss measure will be called \emph{risk-averse}. This dependency limits the extent by which outliers are allowed to violate the requirements. In contrast, formulations in which the resulting design does not depend on a loss measure will be called {\em risk-agnostic}.  

\subsection{Pseudo-distributions}
The infinite set $r_k(\theta, a, E)$ contains the $n_e$ points $r_k(\theta, a, {\mathcal E})$ if ${\mathcal E}\subset E$. These points will be interpreted as realizations of the random variable $r_k(\theta, a, e)$, where $e$ is uniformly distributed in $E$. The distribution associated with such a variable will be referred to as a \emph{pseudo-distribution} in order to distinguish it from the distribution associated with a true random variable. As such, the $n_e$ points $r_k(\theta, a^{(i)},\mathcal{E})$ in (\ref{sta2}) corresponding to a fixed $k$ and a fixed $i$ will be treated as draws from a pseudo-distribution. 

\subsection{Preliminars}
The formulations below use the approximation in the Appendix to represent the {\em Cumulative Distribution Function} (CDF)  of a random variable and its inverse based in a finite number of observations. These approximations will be denoted as $F_{\mathcal Z}$ and $F^{-1}_{\mathcal Z}$ respectively, where ${\mathcal Z}$ is the data sequence. When the random variable is given by a $C^1$ function $f(\theta,a)$ in $\theta$ such approximations are differentiable, thereby making standard gradient-based algorithms applicable. Furthermore, define $0 \leq \alpha_{e,k} \ll 1$ as the fraction of the $n_e$ epistemic scenarios allowed to violate the $k$-th requirement. The analyst might chose $\alpha_{e,k}>0$ in order (i) to relax the demands of such a requirement, (ii) to relax the prescription of an intrinsically uncertain $E$, therefore of ${\mathcal E}$, or (iii) to make the program feasible, i.e., there is no design point for which the requirement is satisfied for all elements of ${\mathcal E}$.
\section{Moment-independent Scenario Programs}\label{RCC} 
In this section we present reformulations of (\ref{sta2}) that do not depend on moments of a response function.  The alternative is studied in Section \ref{MOM}. 

\subsection{Risk-averse formulations}\label{wcs}
Two formulations are presented next. The first formulation eliminates the same epistemic outliers from all the pseudo-distributions associated with ${\mathcal A}$, thereby rendering a comparatively greater $J(\theta^\star)$. Conversely, the second formulation makes the epistemic outliers of each of the $n_a$ pseudo-distributions depend on the corresponding aleatoric scenario, thereby further reducing the objective value. 

\subsubsection{Eliminating a global set of epistemic outliers}
The formulation below identifies and eliminates epistemic outliers by using weights. Consider the weight sequence ${\mathcal W}=\{w^{(j)}\}_{j=1}^{n_e}$, where $0\leq w^{(j)}\leq 1$. The closer $w^{(j)}$ is to zero the less important the $j$-th epistemic scenario becomes. Rather than choosing a fixed set of weights upfront, we will make this choice depend on $\theta$ in order to carry out an optimal selection. Consider the optimization program
\begin{align}
	\label{wc_univ}
	\min_{\theta \in \Theta,\,  \xi \geq 0} & \; J\left(\theta \right) + \rho \sum_{i=1}^{n_a}\xi_{i} \\
	\text{subject to:} & \;\; w_{\theta, k}^{(j)}\;r_k(\theta,a^{(i)},e^{(j)})\leq \xi_{i}, \;  i=1,\ldots n_a,\; j=1,\ldots n_e, \; k=1,\ldots n_r, \nonumber
\end{align}
where $\rho\geq0$ is a penalty parameter, $\xi \in \mathbb{R}^{n_a}$ is a slack variable, and $w_{\theta, k}^{(j)}$ is an element of the weight sequence ${\mathcal W}(\theta,k,\sum_{i=1}^{n_a} \text{sgn}(\xi_i)/n_a,\alpha_{e,k})$ detailed in Appendix. Denote $\theta^\star$ and $\xi^\star$ as the solution to  (\ref{wc_univ}). Depending on the value of $\rho$, some scenarios are allowed to fall onto the failure domain for the purpose of lowering $J(\theta)$, but this action has itself a cost as expressed by $\xi_{i}$: if $\xi_{i} > 0$ the constraint $w_{\theta,k}^{(j)}r_k(\theta,a^{(i)},e^{(j)})\leq 0$ is relaxed to $w_{\theta,k}^{(j)}r_k(\theta,a^{(i)},e^{(j)})\leq \xi_{i}$ in exchange for a cost increase of $\rho\,\xi_{i}$. Hence, (\ref{wc_univ}) is a flexible scheme that enables the analyst to explore various solutions as $\rho$ is varied from zero (no regret for requirement violations) to infinite (infinite regret for requirement violations).  The weights take the value of zero for the epistemic outliers, thereby practically eliminating the corresponding constraints.  Note that the constraints corresponding to a fixed $i$ and a fixed $k$ for $\xi_i=0$ imply that $r_k(\theta,a^{(i)},e^{(j)})> 0$ for no more than $\lfloor n_e\,\alpha_{e,k} \rfloor$ of the scenarios in $\mathcal{E}$.  The epistemic outliers, whose number depends explicitly on the chosen value of $\alpha_{e,k}$, are ${\mathcal O}_e=\left \{e^{(j)}\in{\mathcal E}:  w_{\theta^\star,\, k}^{(j)}=0 \text{ for some }k\in 1,\ldots n_r\right \}$.   

\subsubsection{Eliminating a local set of epistemic outliers}
Consider the optimization program\footnote{This paper uses individual CCs because (i) they are simpler to implement, (ii) the maximally acceptable probability of failure might not be the same for all requirements, and (iii) the proposed framework can readily handle joint CCs by using non-conservative individual CCs (Footnote 1). Furthermore, they will cast reliability requirements in terms of the inverse CDF to prevent convergence to a local optimum.}
\begin{align}
	\label{wc}
	\min_{\theta \in \Theta,\,  \xi \geq 0} & \quad J\left(\theta \right) + \rho \sum_{i=1}^{n_a}\xi_{i} \\
	\text{subject to:} & \quad F_{{\mathcal Y}(\theta,k,\,a^{(i)},\, {\mathcal E})}^{-1} (1-\alpha_{e,k}) \leq \xi_{i}, \;  i=1,\ldots n_a,\; k=1,\ldots n_r, \nonumber
\end{align}
where $\rho\geq0$ is a penalty parameter, $\xi \in \mathbb{R}^{n_a}$ is a slack variable, and
\begin{equation}\label{ze}
{\mathcal Y}(\theta,k,a, {\mathcal E}):= \left\{ r_k(\theta,a, e^{(j)})\right \}_{j=1}^{n_e}. 
\end{equation}
As before, by lowering $\rho$ some scenarios are allowed to fall onto the failure domain for the purpose of reducing $J(\theta^\star)$, but this action has itself a cost as expressed by $\xi_{i}$: if $\xi_{i} > 0$, the constraint $F_{{\mathcal Y}}^{-1} (1-\alpha_{e,k})\leq 0$ is relaxed to $F_{{\mathcal Y}}^{-1} (1-\alpha_{e,k})\leq \xi_{i}$ in exchange for a cost increase of $\rho\,\xi_{i}$. The constraints of (\ref{wc}) corresponding to a fixed $i$ for $\xi_i=0$ imply that $r_k(\theta,a^{(i)},e)>0$ for no more than $\lfloor n_e\, \alpha_{e,k}\rfloor$ of the scenarios in ${\mathcal E}$.  In contrast to (\ref{wc_univ}), however, each pseudo-distribution  can have a different set of epistemic outliers.  Because the removal of such outliers yields the greatest reduction in the value taken by $r_k(\theta,a^{(i)},e)$ for each $i$,  the value of $J(\theta^\star)$ resulting from (\ref{wc}) is often lower than the value of $J(\theta^\star)$ resulting from (\ref{wc_univ}). 


\subsubsection{Discussion}
\vspace{0.2cm}
\noindent {\emph{Remark 1}:} Programs (\ref{wc_univ}) and (\ref{wc}) yield the same design when $\alpha_{e,k}=0$. These formulations are WC because $\xi_i^\star$ depends on the aleatory scenarios at which the requirement functions take on the greatest value. Note that the number of decision variables in both programs grows linearly with $n_a$, thereby limiting their applicability to moderately large datasets and/or convex programs. 

\vspace{0.2cm}
\noindent {\emph{Remark 2}:} 
The epistemic outliers corresponding to the $i$-th aleatory scenario are given by
\begin{equation}
{\mathcal O}_{e}(i)=\left\{e^{(j)}\in{\mathcal E}: r_k(\theta^\star, a^{(i)}, e^{(j)})> F_{{\mathcal Y}(\theta^\star,k,a^{(i)},\, {\mathcal E})}^{-1} (1-\alpha_{e,k}) \text{ for some }k\right \},\nonumber
\end{equation}
whereas the aleatory outliers are given by
\begin{equation}
{\mathcal O}_a=\Big \{ a^{(i)}\in{\mathcal A}: \max_{k=1,\ldots n_r} F_{{\mathcal Y}(\theta^\star,\,k,\, a^{(i)},\, {\mathcal E})}^{-1} (1-\alpha_{e,k})>0\Big \}, \label{outset2}
\end{equation}
i.e., the scenarios for which at least one of the individual failure probability exceeds the acceptable limit. Only the aleatoric outliers contribute to the summation in (\ref{wc_univ}) and (\ref{wc}) by taking on values proportional to the extent of the requirement violation, thereby making such formulations risk-averse.  In this context, the adverse outcome is the violation of a requirement, whereas $\xi_i$ is the loss measure. Hence, the dependency of the penalty term on $\xi$ makes the resulting $\theta^\star$ a risk-averse design. This dependency limits the extent by which outliers are allowed to violate the requirements. 

\vspace{0.2cm}
\noindent {\emph{Remark 3}:} Optimization programs (\ref{wc_univ}) and (\ref{wc}) select outliers optimally. While the number of epistemic outliers depends on the chosen value for $\alpha_{e,k}$, the number of aleatoric outliers depends implicitly on $\rho$. These programs might not have a solution for which all $\xi_{i}^\star$ are zero, thereby making relaxation crucial. Relaxation entails turning a given optimization program into one with either looser constraints or fewer constraints in order to enlarge the feasible set, thereby possibly yielding a lower $J(\theta^\star)$. The feasible set of these programs is 
\begin{equation}
\mathcal{X}_{\text{robust}}=\bigcap_{i\,\in\, {\mathcal I}}\; \mathcal{X}\left (a^{(i)}, {\mathcal E}\setminus  {\mathcal O}_e(i) \right ), \label{fea2}
\end{equation}
where $ {\mathcal I}$ are the indices corresponding to the aleatory inliers. Hence, relaxation is attained by eliminating both aleatory and epistemic outliers. Relaxation enables the analyst to pursue the most robust design of a given class, as well as to identify the aleatoric scenarios for which the specifications cannot be met. 

\subsection{Risk-agnostic formulations} \label{cco}
A key benefit of the above formulations is that they permit finding designs that can only satisfy the requirements for a subset of the aleatoric scenarios without having any prior knowledge about this subset.  However, their risk-averse nature often renders a greater objective value. Conversely, the following formulations do not use a loss measure to carry out the relaxation, thereby becoming practically risk-agnostic. These formulations enable the analyst to prescribe upfront the desired number of the aleatoric and epistemic outliers by setting values for $0\leq \alpha_{a,k}\ll 1$ and $0\leq \alpha_{e,k}\ll 1$ for $k=1,\ldots n_r$, respectively. However, the particular elements of such sets are chosen by the optimization program. Setting overly small values for these parameters might make the optimization program infeasible. Strategies that prevent this outcome are presented below.

Risk-agnostic approaches often yield lower values for $J(\theta^\star)$ than their risk-averse counterparts. This is the result of the epistemic outliers prescribing the upper quantiles of the pseudo-distributions, thereby yielding the greatest expansion of the feasible space.  These formulations also have the advantage of having a number of decision variables that does not increase with $n_a$. This feature is greatly beneficial when the training datasets are moderately large, and the optimization program is non-convex. As before, two CC formulations are presented. The first formulation eliminates the same set of epistemic outliers from all the pseudo-distributions, thereby rendering a design with a comparatively greater objective value.  Conversely, the second formulation eliminates a potentially different set of epistemic outliers from each pseudo-distribution, thereby further reducing $J(\theta^\star)$.

\subsubsection{Eliminating a global set of epistemic outliers}
The CCO program of interest is
\begin{align}
	\label{chancon_sce2}
	\min_{\theta\in \Theta} & \quad J\left(\theta\right)\\
	\text{subject to:} 
	                           & \quad F^{-1}_{{\mathcal Z}(\theta,\,k,\, {\mathcal A}, \,{\mathcal E})}(1-\alpha_{a,k})\leq 0,\; k=1,\ldots n_r,\nonumber
\end{align}
where $0 \leq \alpha_{a,k} \ll 1$ is the fraction of the scenarios in ${\mathcal A}$ allowed to violate the $k$-th requirement, 
\begin{equation}
\label{zb}
{\mathcal Z}(\theta,\,k,\, {\mathcal A}, \,{\mathcal E}):= \left\{ \max_{j=1,\ldots n_e}\,w_{\theta,k}^{(j)}\,r_k(\theta, a^{(i)},e^{(j)})\right \}_{i=1}^{n_a},
\end{equation}
is the sequence of WC requirement values among the epistemic inliers, and $w_{\theta,k}^{(j)}$ is an element of the weight sequence ${\mathcal W}(\theta,k,\alpha_{a,k},\alpha_{e,k})$ prescribed in the Appendix. 
Hence, (\ref{chancon_sce2}) seeks a design that minimizes $J(\theta)$ such that at least $\lfloor n_a\,(1- \alpha_{a,k})\rfloor $ of the aleatoric scenarios satisfy the $k$-th requirement for more than $\lfloor n_e\, (1-\alpha_{e,k})\rfloor$ of the epistemic scenarios. Thus, we obtain $\lfloor n_a\,(1- \alpha_{a,k})\rfloor$ pseudo-distributions with an acceptably low empirical failure probability after the same set of epistemic outliers are eliminated from all of them.

\subsubsection{Eliminating a local set of epistemic outliers}
Consider the program
\begin{align}
	\label{chancon_sce}
	\min_{\theta\in \Theta} & \quad J\left(\theta\right)\\
	\text{subject to:} 
	                           & \quad F^{-1}_{{\mathcal N}(\theta,\,k,\,  \alpha_{e,k}, \,  {\mathcal A},\, {\mathcal E})}(1-\alpha_{a,k})\leq 0,\;k=1,\ldots n_r, \nonumber
\end{align}
where 
\begin{equation}
\label{za}
{\mathcal N}(\theta, k,\alpha,{\mathcal A}, {\mathcal E}):= \left\{ F^{-1}_{\mathcal{Y}(\theta, \,k,\,a^{(i)},\,  {\mathcal E})}(1-\alpha)\right \}_{i=1}^{n_a},
\end{equation}
and $\mathcal{Y}$ is in (\ref{ze}). 
Hence, (\ref{chancon_sce}) seeks a design that minimizes $J(\theta)$ such that no less than $\lfloor n_a (1-\alpha_{a,k})\rfloor $ of the aleatoric scenarios satisfy the $k$-th requirement for no less than $\lfloor n_e (1-\alpha_{e,k})\rfloor$ of the epistemic scenarios. In contrast to (\ref{chancon_sce2}), however, the epistemic outliers are different for each pseudo-distribution.  In particular, the epistemic outliers eliminated from the $i$-th pseudo-distribution attain the greatest $1-\alpha_{e,k}$ quantiles. The constraints in (\ref{chancon_sce2}) and (\ref{chancon_sce}) for $\alpha_{e,k}=0$ are a computationally tractable heuristic for $\mathbb{P}_{a}[{\mathcal F}_k(\theta,E)]\leq \alpha_{a,k}$. 

\subsubsection{Discussion}
\vspace{0.2cm}
\noindent {\emph{Remark 4}:} Strategies to find values for $\alpha_{a}=[\alpha_{a,1},\ldots,\alpha_{a,n_r}]\in{\mathbb R}^{n_r}$ making programs (\ref{chancon_sce2}) and (\ref{chancon_sce}) feasible are presented next.  A simple strategy is to solve (\ref{wc_univ}) or (\ref{wc}) for $\rho\gg 1$ and then make $\alpha_{a,k}$ greater or equal than the fraction of aleatoric scenarios violating the $k$-th requirement. Alternatively, we can solve
\begin{align}
	\label{chancon_scex}
	\min_{\theta\in \Theta,\; 0\leq \alpha_a\leq 1} & \quad \omega^\top \alpha_a\\\
	\text{subject to:} 
	                           & \quad F^{-1}_{{\mathcal Q}(\theta,\,k,\,  \alpha_{e,k}, \,  {\mathcal A},\, {\mathcal E})}(1-\alpha_{a,k})\leq 0,\;k=1,\ldots n_r, \nonumber
\end{align}
where $\omega>0$ and ${\mathcal Q}$ is a place holder for either ${\mathcal Z}$ or ${\mathcal N}$ in (\ref{chancon_sce2}) or (\ref{chancon_sce}) respectively. The resulting $\theta^\star$ minimizes the weighted sum of the individual failure probabilities, whereas the resulting $\alpha^\star_{a}$ is a lower bound to $\alpha_{a}$'s making the corresponding risk-agnostic formulation feasible. 

\vspace{0.2cm}
\noindent {\emph{Remark 5}:} Recall that the aleatoric outliers in (\ref{outset2}) are the scenarios for which at least one requirement is violated by an unacceptably high number of epistemic scenarios. The expansion of the feasible set of (\ref{chancon_sce2}) or (\ref{chancon_sce}) caused by relaxation often yields a lower objective function without impacting most of the data, i.e., the aleatoric inliers corresponding to the optimal design $\theta^\star$ often satisfy the requirements for most of the epistemic outliers.  

\vspace{0.2cm}
\noindent {\emph{Remark 6}:} Designs based on (\ref{chancon_sce2}) and (\ref{chancon_sce}) are practically risk-agnostic because $\theta^\star$ might either not depend on any outlier, i.e., $\theta^\star$ does not depend on any loss measure, or it might depend on a single outlier, i.e., the scenario at which the sequence of requirement values takes the positive value closest to zero. This is a consequence of the number of scenarios prescribing the value of the inverse CDF in \cite{Crespo25}, which is equal to one when $\alpha$ is a multiple of $1/(n-1)$ or two otherwise.

\vspace{0.2cm}
\noindent {\bf Example 1 (data-enclosing set)}:  In this illustrative example we seek to compute a set of minimal size that encloses a number of aleatory points in ${\mathbb R}^2$. The geometry of this set depends on the decision variable $\theta$, comprised by $c\in{\mathbb R}^2$ and $\mu\in{\mathbb R}^+$, and the epistemically uncertain parameter $e \in {\mathbb R}^3$. This is cast by\footnote{The subscript $k$ will be removed from the notation since $n_r=1$.}
\begin{align}
	\min_{c,\; \mu\geq 0} & \quad \pi \mu^2  \label{j1}\\
	\text{ subject to:}& \quad r(\theta, a^{(i)}, e^{(j)})\leq0, \; \text{ for all }a^{(i)}\in \tilde{{\mathcal A}},\; e^{(i)}\in\tilde{{\mathcal E}},\label{con1}
\end{align}
where $r(\theta,a,e)=\|\tilde{c}(\theta,e)-a\|_2^2- \tilde{\mu}(\theta,e)^2$, $\tilde{{\mathcal A}}\subseteq {\mathcal A}$ and $\tilde{{\mathcal E}}\subseteq {\mathcal E}$ are yet to be prescribed, and ${\mathcal E}$ is a subset of $E= [0,1/5] \times [0,2\pi]\times [0,1/5]$.  

The variables $\tilde{c}$ and $\tilde{\mu}$, which prescribe the center and radius of a circle respectively, are given by $\tilde{c}(\theta,e)=c+\mu e_1u$ and $\tilde{\mu}(\theta,e)=\mu(1+\mu e_1 e_3\, c^\top u)$, where $u=[ \cos(e_2), \sin(e_2)]^\top$.  
This prescription of the epistemic uncertainty renders a success domain, ${\mathcal S}(\theta,E)$, given by the intersection of infinitely many circles having different centers and radii. Each of these circles correspond to a different realization of the epistemic uncertainty.  We can imagine that the analyst chooses a nominal circle with center $c$ and radius $\mu$ by setting the value of $\theta$, but a different circle is actually implemented. The circle implemented is an unknown element of a set of infinitely many circles depending on both $\theta$ and $E$. As such, the goal of the designer is to prescribe a value of $\theta$ such that all (or most of) the points in ${\mathcal A}$ are tightly enclosed by ${\mathcal S}(\theta,E)$, i.e., by the intersection of all (or most of) possible perturbations of the nominal circle.  
Salient features of designs resulting from the above formulations are presented next. Figures of merit for such designs are listed in Table \ref{tab}.
\begin{table}[h!]
\scriptsize
\setlength{\tabcolsep}{2pt}
\caption{\tiny{The symbols at the top of the table were defined above, whereas those in the midsection result from a Monte Carlo analysis with $n_a'=10000$ and $n_e'=200$ scenarios. They include the range of failure probabilities $a$ in (\ref{pfrange}); the range $b$ in (\ref{bb}), and the $95\%$ confidence interval of the probability of satisfying the robustness specifications $d$ in (\ref{mixed_ci}) for $\alpha'_a=0$, $\alpha'_e=0$, and $P^{\text{max}}=0.01$. At the bottom, we list the number of support scenarios, $n_s$, the number of set violations, $n_v$, the set-complexity, $s_E$, and the scenario bound $\overline{\epsilon}(s_E)$ for $\beta=\num{1e-4}$.}}
\centering
 \label{tab}
 \begin{tabular}{|c| c c c c c c |c c |c c|cc|}
 \hline 
    &  $\theta_1^\star$  & $\theta_2^\star$ &  $\theta_3^\star$  & $\theta_4^\star$ & $\theta_5^\star$  &$\theta_6^\star$  &$\theta_7^\star$  &$\theta_8^\star$  &$\theta_9^\star$  & $\theta_{10}^\star$   &$\theta_{11}^\star$  &$\theta_{12}^\star$  \\ [0.5ex]
\hline
\text{Program}                                                               &  (\ref{wc})         &   (\ref{wc})           & (\ref{wc_univ})  &(\ref{wc})               &(\ref{chancon_sce2})   &(\ref{chancon_sce})   &(\ref{chancon_sce})    &(\ref{chancon_sce})    &(\ref{chancon_sce4})  &(\ref{chancon_sce4})          &  SD       &  SD \\[0.5ex] 
\textcolor{red} {\text{Objective}}                                    &  (\ref{j1})           &  (\ref{j1})              & (\ref{j1})            &(\ref{j1})                  &(\ref{j1})                   & (\ref{j1})                        &(\ref{j1})                      &(\ref{j1})    &(\ref{mea})                &(\ref{mea})                         &  (\ref{j1})  &  (\ref{j1})   \\[0.5ex] 
$n_a$                                                                                 &  50                    &   50                      & 50                    &50                          &50                             &50                                 &100                             &5000   &50                               &50                                         &  80              &  80 \\[0.5ex] 
$n_e$                                                                                  &  50                    &   50                      & 50                    &50                           &50                            &50                                &100                             &100      &50                               &50                                         & 64              &  1\\[0.5ex] 
$|\mathcal{O}_a|$                                                               &  0                      &   2                        &2                        &2                            &2                              &2                                 &0                                 &2            &0                                 &2                                            & NA           & NA \\[0.5ex] 
$\alpha_e$                                                                          &  0                      &   0                        &2/49                   &2/49                        &2/49                        &2/49                             &0                                 &0             &0                                 &2/49                                        & 0             & 0\\[0.5ex] 
$J$                                                                                      &  137.9               &  78.6                   &76.7                    &75.4                        &74.5                        &68.4                             &139.2                          &255.9       &47.5                          &34.6                                         & 245.7      & 128.8 \\[0.5ex] 
\hline
$\underline{a}/\num{1e-3}$                                                & 1.2                    & 3.9                     &2.4                      &1.1                         &2.3                          &4.1                                 &1.3                              &0          &1.2                           &1.0                                              &   0.2          & 0.1\\[0.5ex] 
$\overline{a}/\num{1e-3}$                                                  &  7.2                   &  33.8                   &47.3                    &66.5                       &42.9                        &57.1                                &6.9                              &1.6         &7.0                            &32.9                                            & 1        & 24.4\\[0.5ex] 
$\underline{b}/\num{1e-3}$                                                &  0.5               &  2.7                      &1.5                    &0.5                          &1.4                         &2.9                                  &0.6                              &0            &0.5                              &0.4                                           & 0             &  0\\[0.5ex] 
$\overline{b}/\num{1e-3}$                                                   &  8.9                 &  37.4                   &51.6                    &71.5                       &47.0                        &61.7                                &8.6                              &2.5       &8.8                            &36.4                                             & 2.5         & 27.5\\[0.5ex] 
$\underline{d}/\num{1e-3}$                                                 &0                       & 732                    &657                     &443                        &647                        &891                                 &0                                 &0           &0                                &182                                                   & 0             & 5.5\\[0.5ex] 
$\overline{d}/\num{1e-3}$                                                   &18.2                  & 849                    &786                     &586                         &777                       &965                                  &18.2                            &18.2      &18.2                          &305                                                   & 18.2        & 50.4\\[0.5ex] 
\hline
$n_s$                                                                                    &  2                       & 4                        &2                       &10                         &10                       &7                                     &2                                 &4           &50                                &50                                                   &NA        & NA\\[0.5ex] 
$n_v$                                                                                    &  2                       &  3                       &7                       &11                         &6                         &8                                     &1                                 &3           &50                                &50                                                   &NA        & NA\\[0.5ex] 
$s_E$                                                                                    &  2                       & 4                        &8                       &18                         &13                       &10                                    &2                                 &4           &50                                &50                                                   &NA        & NA\\[0.5ex] 
$\overline{\epsilon}(s_E)$                                                     & \num{0.303}  & \num{0.369}           &\num{0.477}      &\num{0.687}         &\num{0.590}        &\num{0.525}                      &\num{0.164}           &\num{0.0044}      &1                            &1                                                   &NA        & NA\\[0.5ex] 
\hline
\end{tabular}
\end{table}
           
\begin{figure*}[h!]
    \centering
    \begin{subfigure}[b]{0.25\textwidth}
        \centering
        \centerline{\includegraphics[trim={0cm 1.5cm 3.5cm 3.35cm},clip,width=3.2\textwidth]{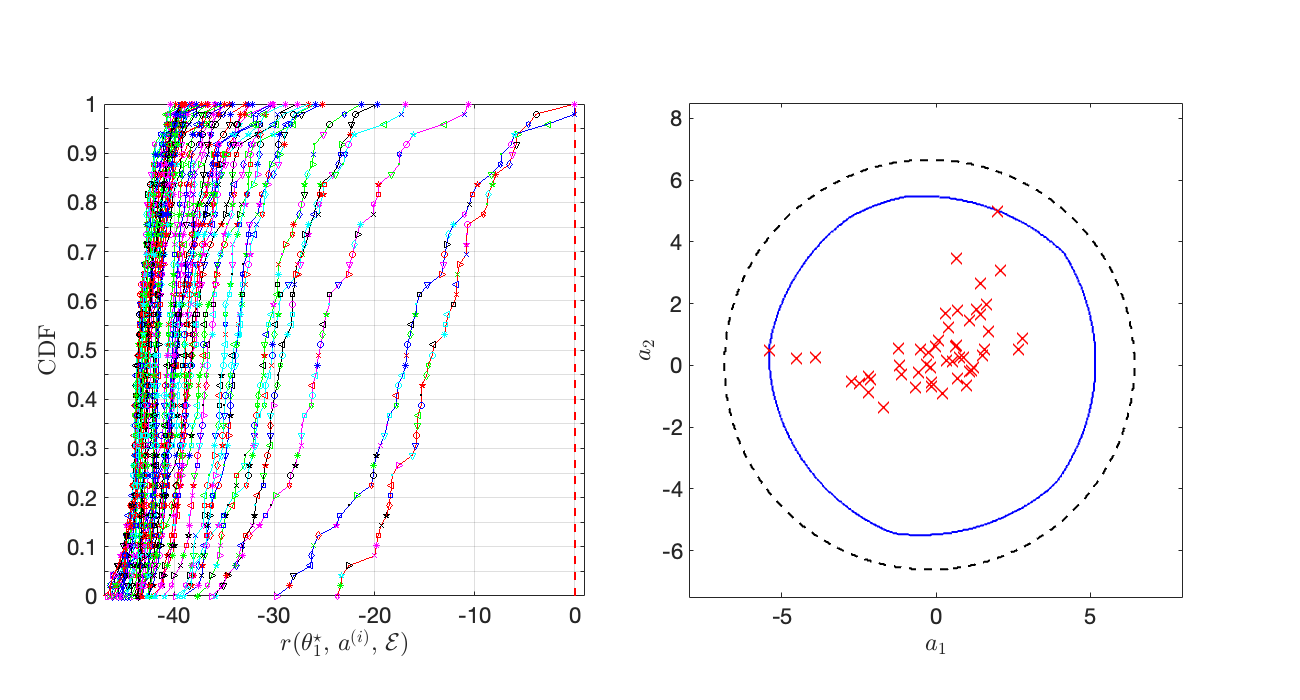}}
    \end{subfigure}
    \vfill
    \begin{subfigure}[b]{0.25\textwidth}
        \centering
        \centerline{\includegraphics[trim={0cm 1.5cm 4cm 3.35cm},clip,width=3.2\textwidth]{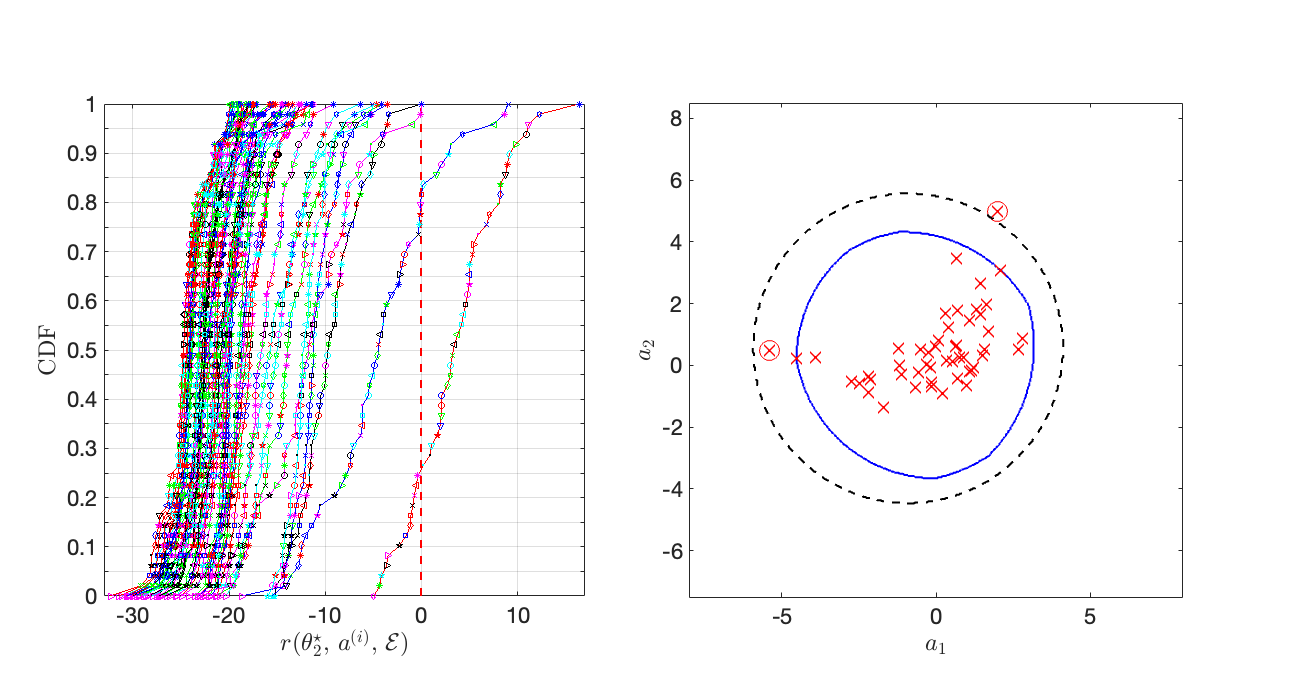}}
    \end{subfigure}
    \hfill
    \caption{Pseudo-distributions of $r(\theta,a^{(i)},{\mathcal E})$  for $i=1,\ldots 50$ (left) and aleatory space (right) for $\theta_1^\star$ (top) and $\theta_2^\star$ (bottom). Left plots:  each epistemic scenario is marked with a distinctive symbol, with those being neglected (if any) being encircled. Right plots: the nominal circle is shown with a dashed black line, and the boundary of the success domain is shown as a solid blue line. The aleatory scenarios are marked with a ``{\color{red} $\times$}'', where the aleatory outliers are marked with ``{\color{red} $\circ$}'' (if any).}
    \label{fig_d12}
\end{figure*}

We start by assuming small training datasets having $n_a=50$ and $n_e=50$ scenarios. The risk-averse formulation (\ref{wc}) with $\alpha_e=0$ was used first to synthesize $\theta_1^\star$ and $\theta_2^\star$, which ignore $|{\mathcal O}_a|=0$ and $|{\mathcal O}_a|=2$ outliers respectively. The left subplots in Figure \ref{fig_d12} show the $n_a$ pseudo-distributions, whereas the right subplots show the aleatoric scenarios (red crosses), the nominal circle (dashed-line) and the boundary of the success domain $\mathcal{S}(\theta,{\mathcal E})$ for $\alpha_e=0$ (solid-line). The aleatoric outliers, whose distributions satisfy $F_{r(\theta,a^{(i)},{\mathcal E})}(0)<1$, are shown as red circles. The elimination of two aleatoric outliers reduced the objective value by $43\%$ in exchange for a loss in robustness (details given in Section \ref{ua}).   

\begin{figure*}[ht!]
    \centering
    \begin{subfigure}[b]{0.25\textwidth}
        \centering
        \centerline{\includegraphics[trim={0cm 1.5cm 3.5cm 3.35cm},clip,width=3.2\textwidth]{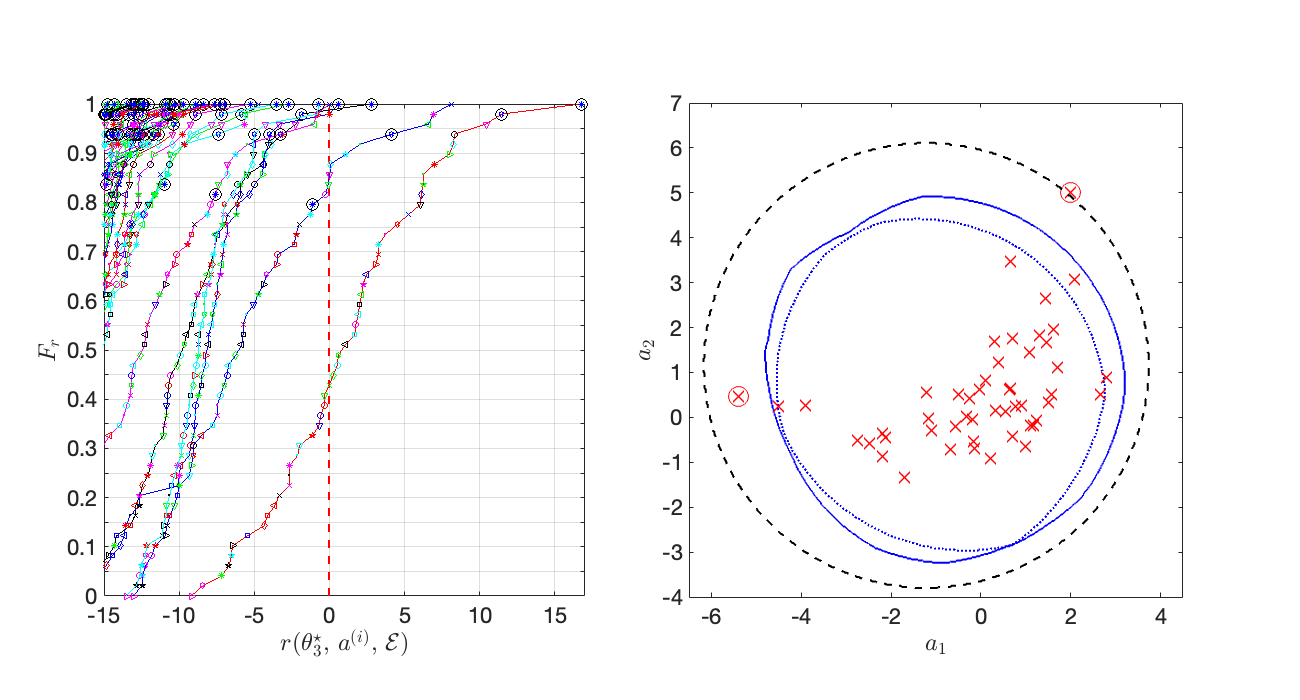}}
    \end{subfigure}
    \vfill
    \begin{subfigure}[b]{0.25\textwidth}
        \centering
        \centerline{\includegraphics[trim={0cm 1.5cm 4cm 3.35cm},clip,width=3.2\textwidth]{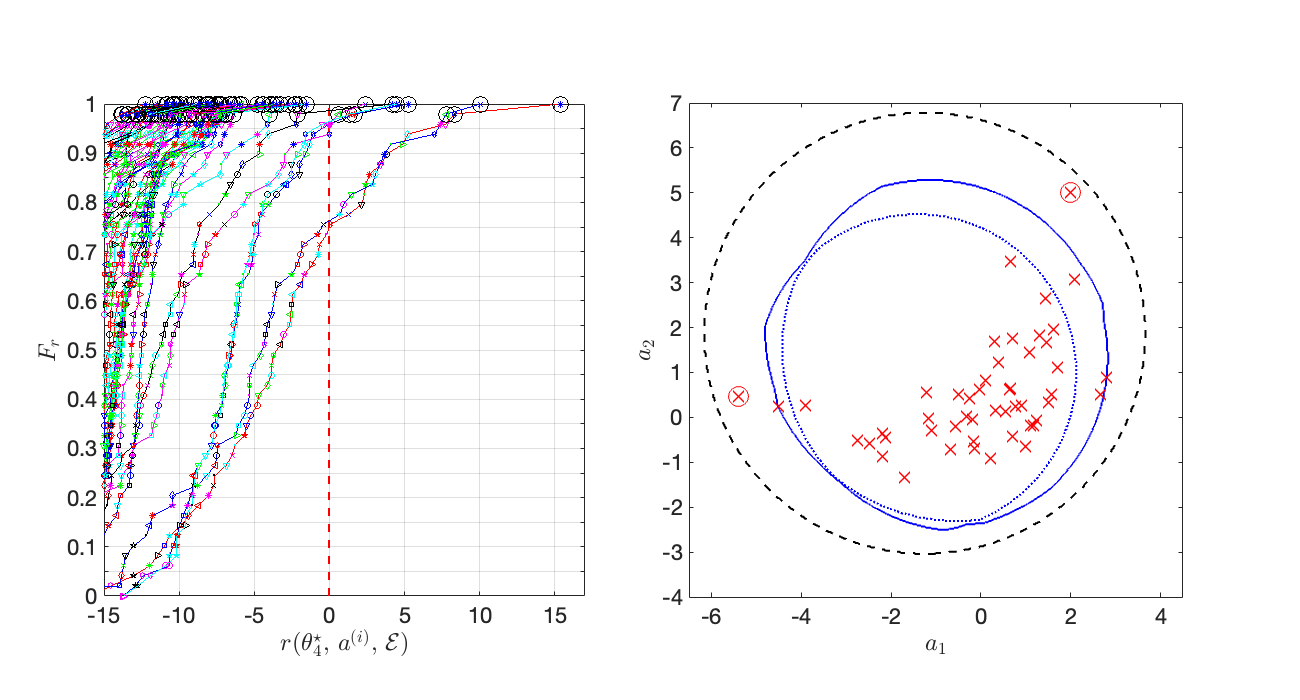}}
    \end{subfigure}
    \hfill
    \caption{Pseudo-distributions of $r(\theta,a^{(i)},{\mathcal E})$ for $i=1,\ldots 50$ (left) and aleatory space (right) for $\theta_3^\star$ (top) and $\theta_4^\star$ (bottom). Previous conventions apply. The success domain corresponding to all $n_e$ epistemic scenarios is shown as a dotted blue line on the right subplots.}
    \label{fig_d34}
\end{figure*}

The risk-averse formulations (\ref{wc_univ}) and (\ref{wc}) were used to synthesize $\theta_3^\star$ and $\theta_4^\star$ after ignoring two epistemic outliers, i.e., $\alpha_e=2/49$. Figure \ref{fig_d34} shows the corresponding pseudo-distributions, the nominal circle, and the boundaries of the success domain for $\alpha_e=0$ (solid-line) and $\alpha_e=2/29$ (dotted-line). Note that more than two of the pseudo-distributions cross the $r=0$ threshold, but only two of them correspond to aleatoric outliers. The aleatoric outliers are the scenarios for which the empirical failure probability exceeds the admissible limit, i.e., $1-F_{r(\theta,a^{(i)},{\mathcal E})}(0)>2/49$. The key difference between $\theta_3^\star$ and $\theta_4^\star$ is the manner by which the epistemic outliers are chosen. Whereas $\theta_3^\star$ chooses the same two epistemic outliers for all $n_a=50$ pseudo-distributions, $\theta_4^\star$ chooses the two WC outliers for each of such distributions. These scenarios are shown as black circles on the left subplots. The elimination of outliers expands the success domain and reduces the objective value $J(\theta)$, with $\theta_4^\star$ attaining the greatest improvement. This expansion is the area between the solid-line and the dotted-line. Note that this area is more than $100\times(2/49)=4.1\%$ of the area of both success domains. 
CCed designs render a lower objective by ignoring the epistemic scenarios falling onto the greatest quantiles of the pseudo-distributions, i.e., the right tail. The further into such a tail the epistemic outliers are, the greater the reduction. The non-trivial dependency of these tails on $\theta$ makes the upfront selection of outliers difficult. Note that two (top) and four (bottom) of the pseudo-distributions corresponding to aleatoric inliers attain positive failure probabilities. 

\begin{figure*}[h!]
    \centering
    \begin{subfigure}[b]{0.23\textwidth}
        \centering
        \centerline{\includegraphics[trim={0cm 1.5cm 3.5cm 3.35cm},clip,width=3.2\textwidth]{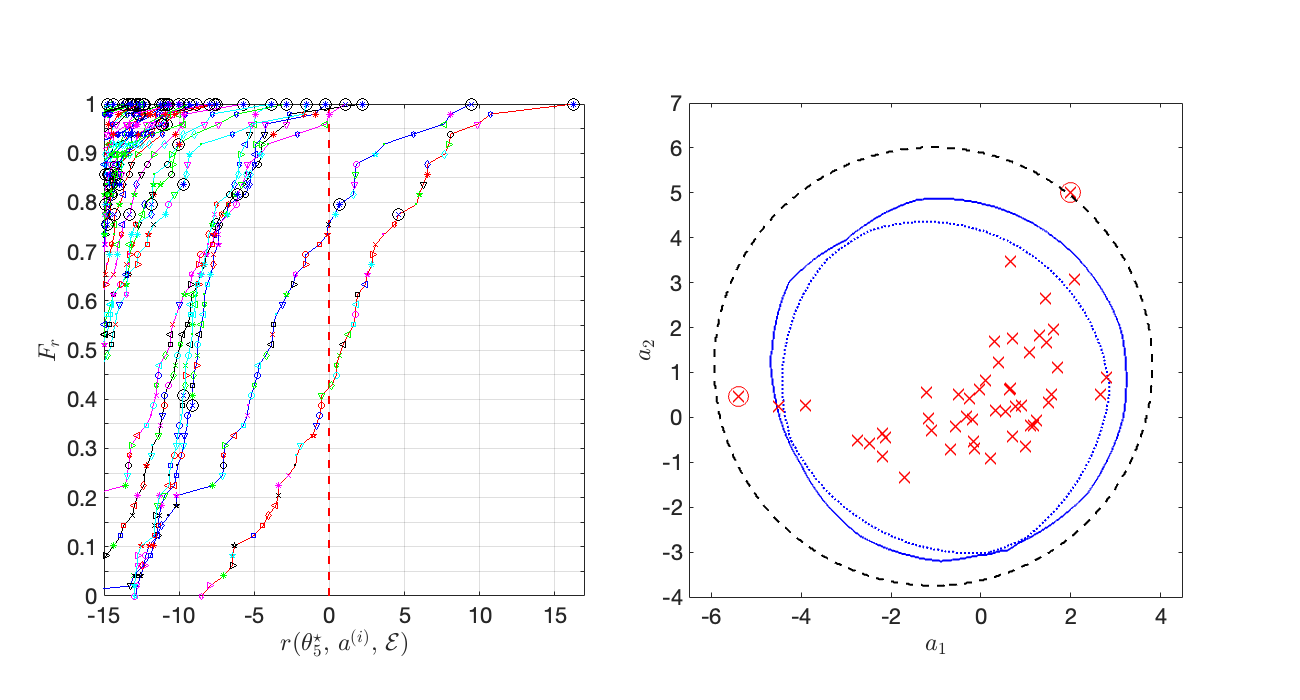}}
    \end{subfigure}
    \vfill
    \begin{subfigure}[b]{0.23\textwidth}
        \centering
        \centerline{\includegraphics[trim={0cm 1.5cm 4cm 3.35cm},clip,width=3.2\textwidth]{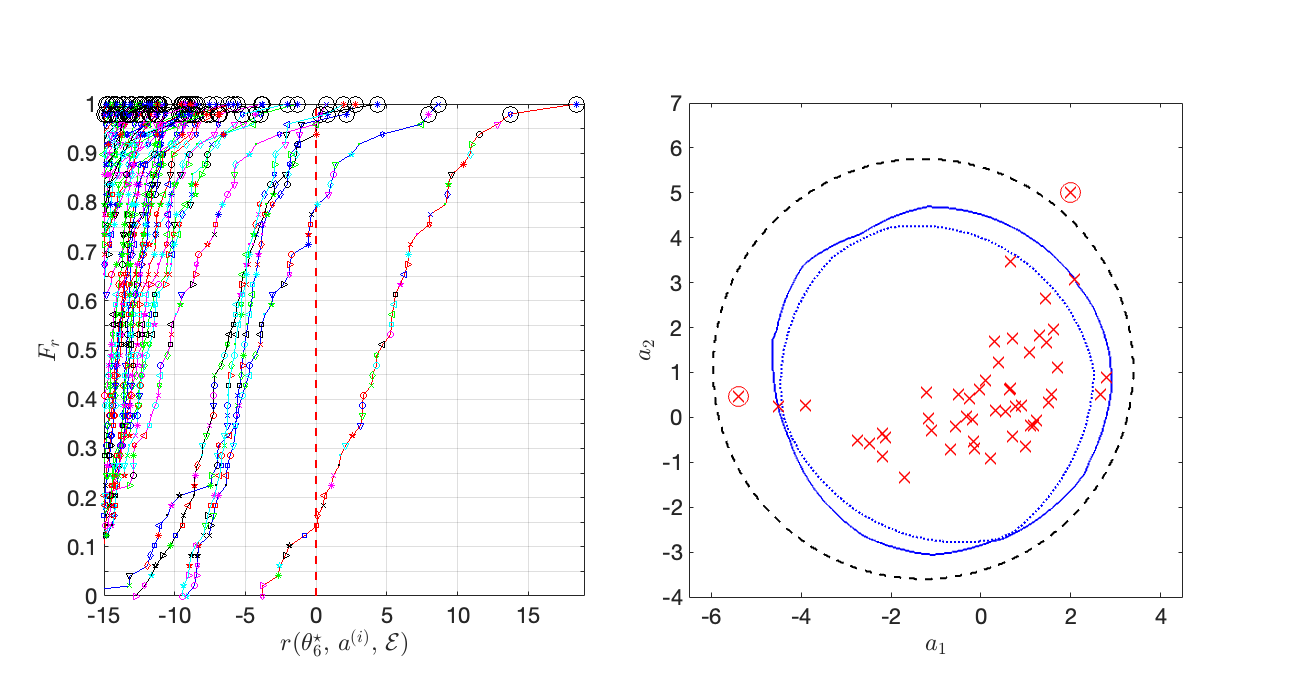}}
    \end{subfigure}
    \hfill
    \caption{Pseudo-distributions of $r(\theta, a^{(i)},{\mathcal E})$ for $i=1,\ldots 50$ (left) and aleatory space (right) for $\theta_5^\star$ (top) and $\theta_6^\star$ (bottom). Previous conventions apply.}
    \label{fig_d56}
\end{figure*}

The risk-agnostic formulations (\ref{chancon_sce2}) and (\ref{chancon_sce}) were used to synthesize $\theta_5^\star$ and $\theta_6^\star$ respectively after ignoring two epistemic outliers. We expect $\theta_5^\star$ and $\theta_6^\star$ to yield a lower objective value than $\theta_3^\star$ and $\theta_4^\star$ since the selection of outliers do not depend on any loss measure. These reductions are 2.9\% and 9.2\% relative to their risk-averse counterparts.  Figure \ref{fig_d56} shows the corresponding pseudo-distributions and the aleatoric space. Note that two (top) and four (bottom) aleatoric scenarios fall onto the expanded portion of the success domain created by ignoring epistemic outliers. The risk-agnostic nature of the formulation makes the aleatoric outliers be further away from the success domain. Further notice that the success domains for $\theta_3^\star$, $\theta_4^\star$, $\theta_5^\star$ and $\theta_6^\star$ differ significantly even though they were trained by the same datasets for the same number of outliers. 

\begin{figure*}[h!]
    \centering
    \begin{subfigure}[b]{0.23\textwidth}
        \centering
        \centerline{\includegraphics[trim={0cm 1.5cm 3.5cm 3.35cm},clip,width=3.2\textwidth]{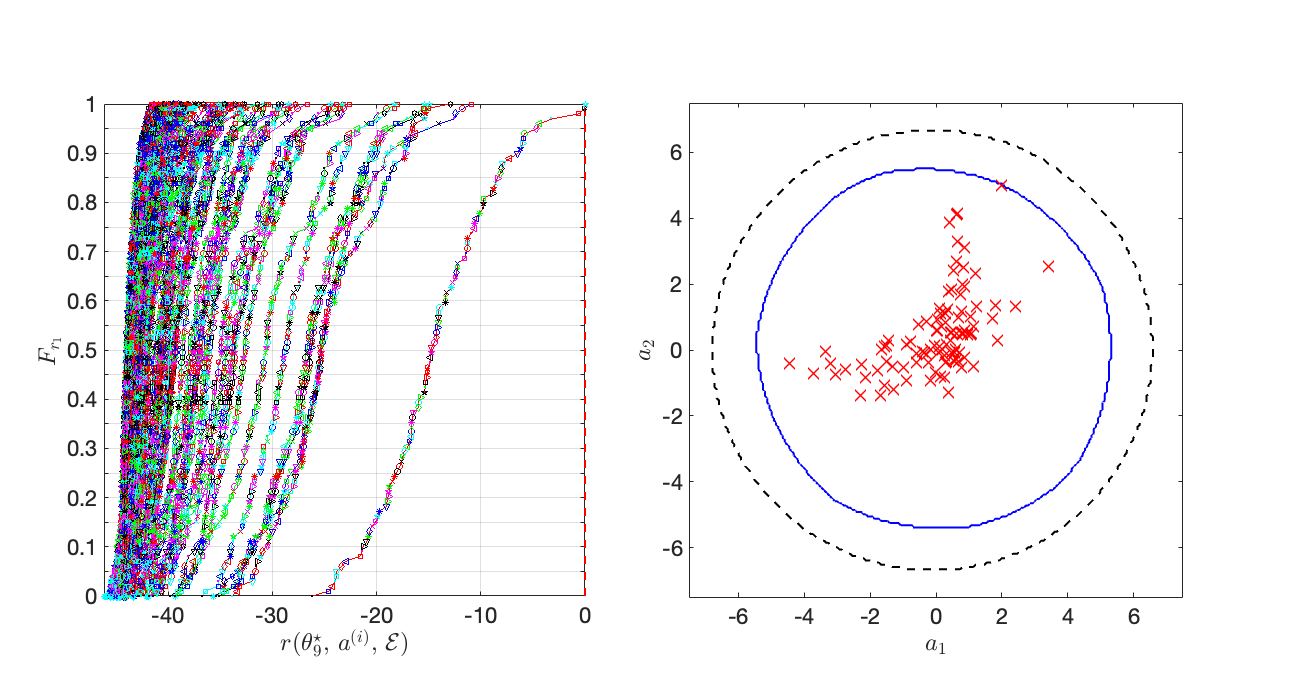}}
    \end{subfigure}
    \vfill
    \begin{subfigure}[b]{0.23\textwidth}
        \centering
        \centerline{\includegraphics[trim={0cm 1.5cm 4cm 3.35cm},clip,width=3.2\textwidth]{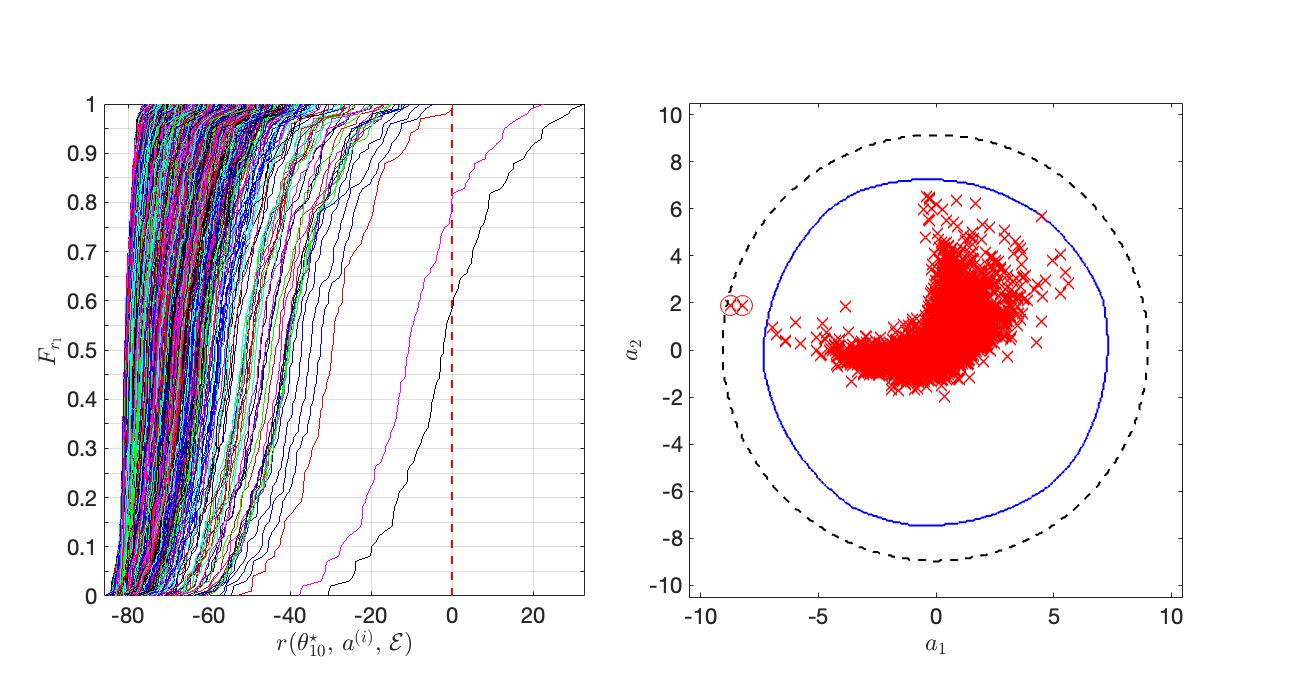}}
    \end{subfigure}
    \hfill
    \caption{Distributions of $r(\theta, a^{(i)},{\mathcal E})$ for $i=1,\ldots n_a$ (left) and aleatory space (right) for $\theta_7^\star$ (top) and $\theta_{8}^\star$ (bottom). Previous conventions apply.}
    \label{fig_d78b}
\end{figure*}

Finally, we present results for designs $\theta_7^\star$ and $\theta_{8}^\star$, which were trained using $n_a=100$ and $n_a=5000$ aleatoric scenarios respectively, and $n_e=100$ epistemic scenarios. The search for these designs entails evaluating the requirement function $n_a \times n_e$ times for each function evaluation of the optimizer. As expected, the greater $n_a\times n_e$, the more accurate the failure probability estimates driving the search for $\theta^\star$ become, and the higher the computational cost. The cost of performing these many evaluations is acceptable in this toy example, but not in most engineering applications. In such cases, the sequential algorithm proposed in Section \ref{sd} should be used. Figure \ref{fig_d78b} shows the corresponding pseudo-distributions and aleatoric spaces. Note that the success domains are larger than those obtained before, and that the outliers are now chosen from the long density tail in the second quadrant.  

\section{Moment-dependent Scenario Programs} \label{MOM}
Designs that minimize a moment of a function depending on $\theta$, $a$ and $e$ are pursued next. This is accomplished by using $J=\lambda$ and appending ${\mathbb M}[{\mathcal H}]<\lambda$ to (\ref{sta2}), where ${\mathbb M}$ is an empirical moment of the dataset 
\begin{equation}\label{ff}
{\mathcal H}(\theta,{\mathcal A}, {\mathcal E},\alpha_e):=\left\{h^{(i)}\right\}_{i=1}^{n_a}, 
\end{equation}
$h^{(i)}:=F^{-1}_{h(\theta,\,a^{(i)},\,\mathcal{E})}(1-\alpha_e)$ is a quantile, $h(\theta,a,e)$ is a response function, $\alpha_e$ is the fraction of the epistemic scenarios in ${\mathcal E}$ to be ignored, and $\lambda$ is an additional decision variable. Therefore, we want to minimize a moment of the $1-\alpha_e$ quantile of all $n_a$ pseudo-distributions.

A key objective of the forthcoming formulations is to consistently eliminate the same set of aleatoric outliers from both the moment calculation and the CCs. 
The selection of outliers will be carried out by using $\theta$ dependent weights. To this end, denote as ${\mathbb M}\,[{\mathcal H}, {\mathcal W}(\theta)]$ an empirical moment of ${\mathcal H}$ according to the weights ${\mathcal W}(\theta)=\{w_\theta^{(i)}\}_{i=1}^{n_a}$, where $0\leq w_\theta^{(i)}\leq 1$. For instance, the mean is
\begin{equation}
{\mathbb M}\,[ {\mathcal H}, {\mathcal W}(\theta)]=\frac{1}{\sum_{i=1}^{n_a} w_\theta^{(i)}}\sum_{i=1}^{n_a} \; h^{(i)} w_\theta^{(i)}. \label{mea}
\end{equation}
Once the search for $\theta^*$ converges, the weights will take values near zero or one. Therefore, the weighted mean in (\ref{mea}) becomes the unweighted empirical mean for an optimally chosen subset of aleatoric inliers. As before, the epistemic outliers can be chosen either globally or locally. The following developments focus on the latter case.
\subsection{Risk-averse formulation}
An extension of the scenario program (\ref{wc}) is
\begin{align}
	\label{mom_chancon_sce}
	\min_{\theta\in \Theta,\; \lambda,\; \xi\geq0} & \quad \lambda+\rho\sum_{i=1}^{n_a} \xi_{i}\\
	\text{subject to:} & \quad F_{{\mathcal Y}(\theta,\,k,\, a^{(i)},\, {\mathcal E})}^{-1} (1-\alpha_{e,k}) \leq \xi_{i}, \;  i=1,\ldots n_a,\;  k=1,\ldots n_r, \nonumber \\
	                           & \quad   \ell(\theta, {\mathcal A}, {\mathcal E}, \alpha_e,{\mathcal W},\lambda) \leq 0, \nonumber
\end{align}
where ${\mathcal W}=\{\exp(-\kappa\, \xi_i)\}_{i=1}^{n_a}$ for $\kappa\gg1$, and
\begin{equation}
\ell(\theta, {\mathcal A}, {\mathcal E}, \alpha_e,{\mathcal W},\lambda):={\mathbb M}\left[ {\mathcal H}(\theta, {\mathcal A}, {\mathcal E},\alpha_e),\, {\mathcal W}\right] - \lambda. \label{l}
\end{equation}
Denote as $\theta^\star$, $\lambda^\star$ and $\xi^\star$ the solution to (\ref{mom_chancon_sce}). Note that ${\mathcal W}$ evaluated at $\xi^\star$ assigns a weight of one to the aleatoric inliers and a value near zero to the aleatoric outliers. Hence, (\ref{mom_chancon_sce}) seeks a design $\theta^\star$ that minimizes the sum of the mean of the $1-\alpha_e$ quantiles of the response function for the inliers and the penalty term  considered earlier.  Note that the decision variable $\xi$ allows for a consistent elimination of outliers.

\subsection{Risk-agnostic formulation} \label{rff}
Note that the decision variable $\xi$, used to consistently eliminate outliers, is no longer available. A risk-agnostic extension of (\ref{chancon_sce}) is presented next. For any $u\in\{1,\ldots n_a\}$, define $g \in \mathbb{R}^{n_r+1}$ as 
\begin{align}
g(u,\theta,{\mathcal A},{\mathcal E}, {\mathcal W},\lambda)& :=\left[\ell(\theta, {\mathcal A}, {\mathcal E}, \alpha_e,{\mathcal W},\lambda),\; \left \{F^{-1}_{\mathcal{Y}(\theta, k,a^{(u)}, {\mathcal E})}(1-\alpha_{e,k})\right\}_{k=1}^{n_r}\right],
\end{align}
where 
$w_\theta^{(i)}=I[d^{(i)}\leq {\mathcal H}(u)]$ are the weights, $I[\cdot]$ is the indicator function, 
and ${\mathcal H}(u)$ is the $u$-th order statistic of ${\mathcal H}$.

An aleatoric scenario $u$ for which $g(u,\theta,{\mathcal A},{\mathcal E},{\mathcal W},\lambda)\leq 0$ satisfies two properties. First, the moment of the response for the lowest $u$ values in ${\mathcal H}$ does not exceed $\lambda$. Second, $u$ is an inlier since all the pseudo-distributions corresponding to the requirements have an admissibly large probability of success. This sets the stage for the optimization program
\begin{align}
	\label{chancon_sce4}
	\min_{\theta\in \Theta,\, \lambda} & \quad \lambda\\
	\text{subject to:} & \quad F^{-1}_{ {\mathcal G}(\theta,\, {\mathcal A},\, {\mathcal E},\,\mathcal{W},\, \lambda)}(1-\alpha_a)\leq 0, \nonumber 
\end{align}
where ${\mathcal G}(\theta, {\mathcal A}, {\mathcal E},\mathcal{W}, \lambda)=\{\max \, g(i,\theta,{\mathcal A},{\mathcal E},{\mathcal W}, \lambda) \}_{i=1}^{n_a}$ and $0\leq \alpha_a\ll 1$ is the greatest fraction of the $n_a$ aleatory scenarios allowed to either violate a requirement or to be excluded from the moment calculation. Denote as $ \theta^\star$ and $\lambda^\star$ the solution to (\ref{chancon_sce4}).
Hence, the design $\theta^\star$ minimizes the sample moment of the response for $\floor{n_a(1-\alpha_a)}$ of the scenarios in ${\mathcal A}$ and $\floor{n_e (1- \alpha_e)}$ of the scenarios in ${\mathcal E}$ while satisfying the requirements with an acceptably large probability of success.  The $\max$ operator in ${\mathcal G}$ ensures that the outliers are eliminated consistently. As before, this formulation often leads to a lower optimal moment than its risk-averse counterpart.

\vspace{0.2cm}
\noindent {\bf Example 2 (data-enclosing set)}: Next we revisit Example 1 but use the empirical mean of $h(\theta,a,e)=\tilde{\mu}(c,\mu,e)^2+\| a-\tilde{c}(c,\mu,e)\|_2$ as the objective function. This objective is a measure of the tightness of the data enclosure.

\begin{figure}[htb]
 \centering
 \includegraphics[trim={1.8cm 2cm 3.5cm 3cm},clip,width=5in]{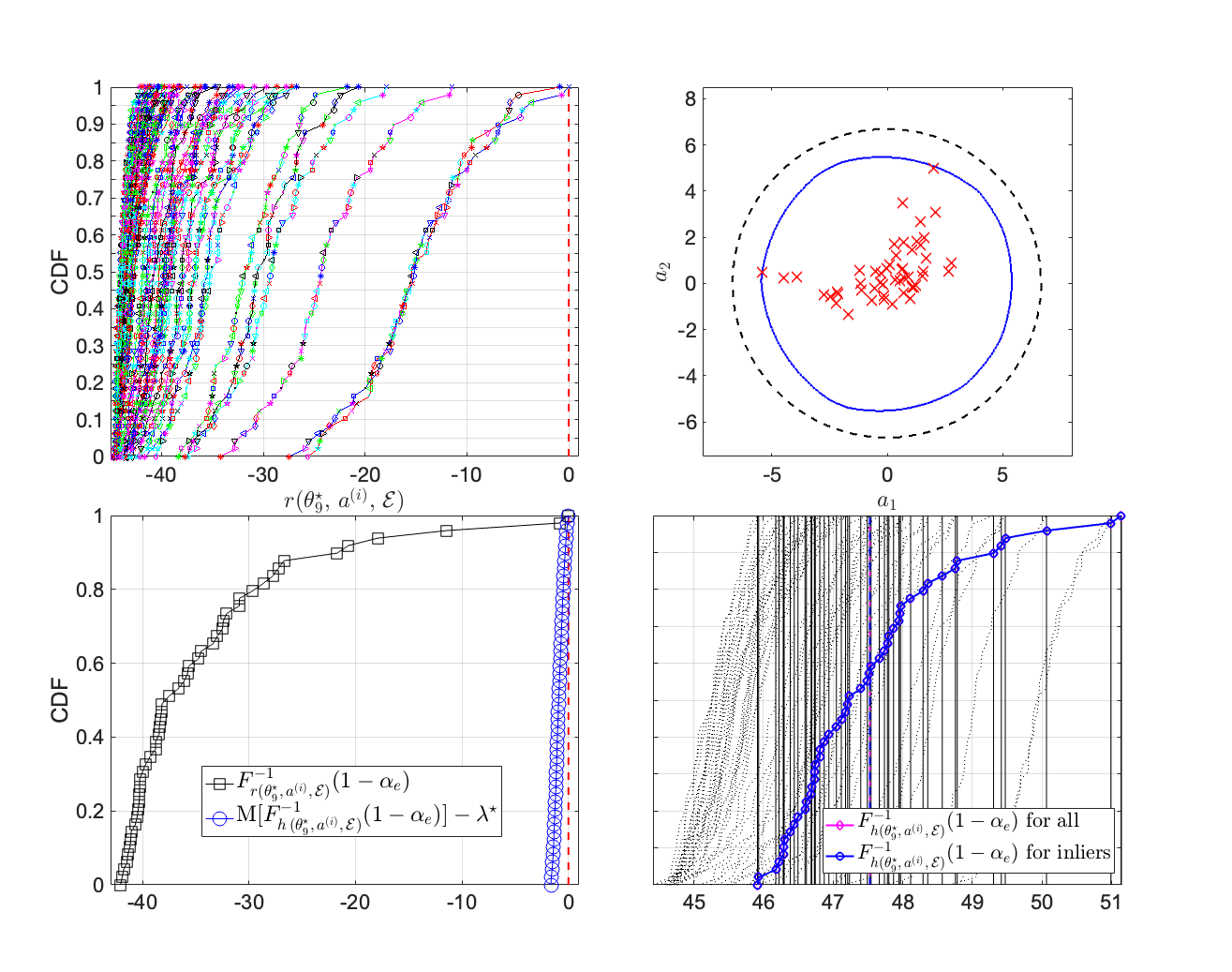}
 \caption{Pseudo-distributions of $r(\theta_9^\star, a^{(i)},{\mathcal E})$ for $i=1,\ldots 50$ (top left), the aleatory space (top right), the distributions of $r$ and of the moment exceedance for $\alpha_e=0$ (bottom left), and the distributions of $h$ for all aleatory scenarios and for the inliers only (bottom right).  The two vertical dashed lines in the bottom-right subplot are the corresponding color-coded moments. Previous conventions apply.}
 \label{fig_d9}
\end{figure}
The risk-agnostic formulation (\ref{chancon_sce4}) was used to synthesize $\theta_9^\star$ and $\theta_{10}^\star$ for a large value of $\rho$.  The top of Figure \ref{fig_d9} shows the pseudo-distributions and the aleatory space for $\theta_9^\star$. The bottom-left subplot shows the empirical distributions of the  one-quantiles, i.e., $\alpha_{e}=0$, of $r(\theta_9^\star)$ and $h(\theta_9^\star)-\lambda_9^\star$. Each point on these distributions correspond to an aleatory scenario. Since none of these distributions cross zero, there are no aleatoric outliers, and all scenarios are used to compute the mean. The bottom-right subplot shows the empirical distribution of the quantiles of the response for both the full set of aleatoric scenarios (shown in magenta), and for the set of aleatoric inliers (shown in blue). These distributions coincide since $|{\mathcal O}_a|=0$. The points comprising these distributions, marked with circles/diamonds, are the $1$-quantile of the pseudo-distributions of the response. These pseudo-distributions are shown as dotted lines. More importantly, the mean of the response corresponding to the blue and magenta distributions are shown as vertical dashed lines with the same color. Both such lines coincide since $|{\mathcal O}_a|=0$.

\textcolor{red} {
\begin{figure}[htb]
 \centering
 \includegraphics[trim={2cm 2cm 4cm 3cm},clip,width=5in]{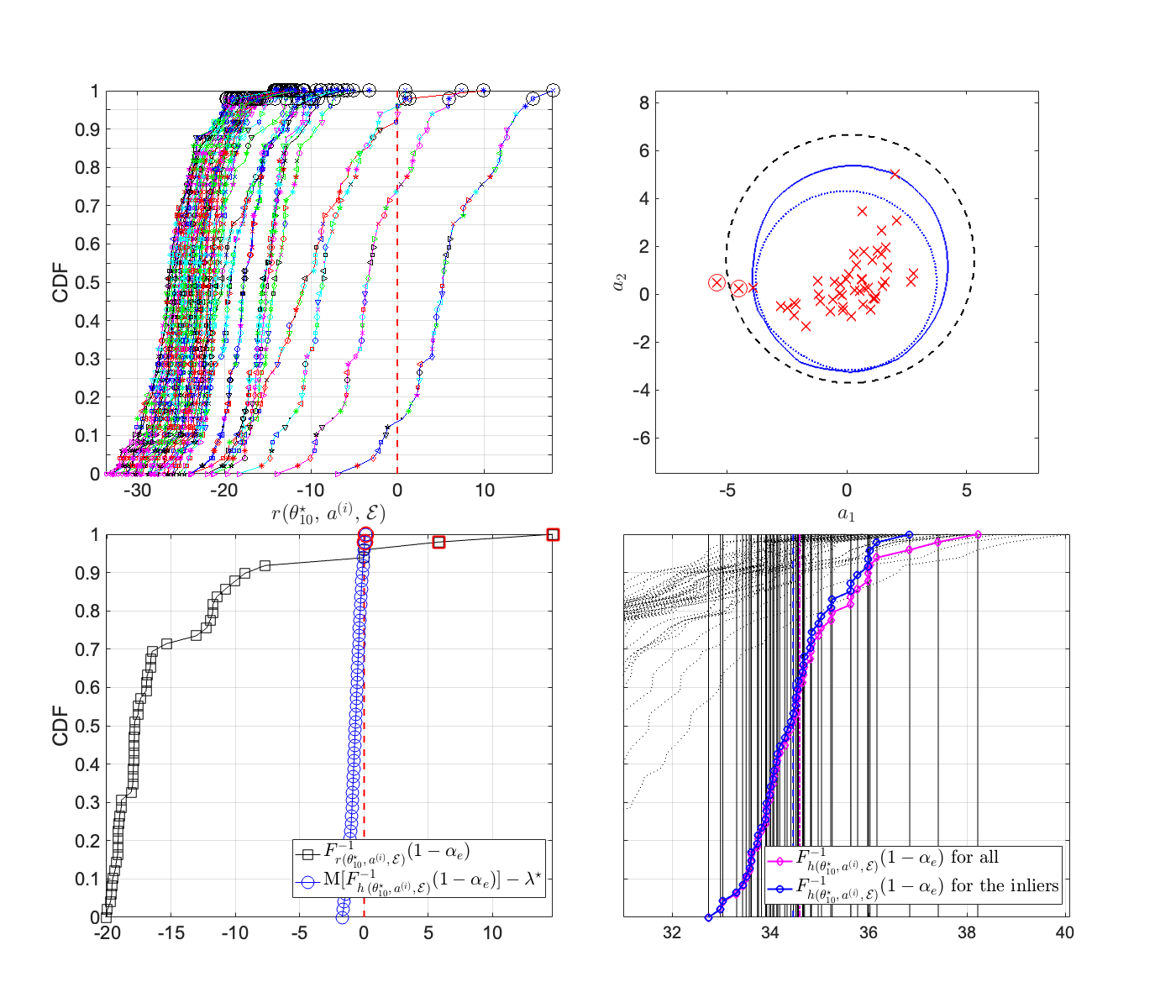}
 \caption{Same information shown in the previous figure but for $\theta_{10}^\star$ and $\alpha_e=2/49$. The requirement and response values corresponding to the outliers are shown with red markers in the bottom-left subplot.}
 \label{fig_d10}
\end{figure}
}
Figure \ref{fig_d10} shows the same information for $\theta_{10}^\star$, which ignores two aleatoric and two epistemic outliers. The elimination of outliers yielded a reduction in the mean of 27.1\% (from 47.5 to 34.6). This performance improvement is the result of ignoring two aleatoric outliers (those with a failure probability greater than $\alpha_{e}=2/49$), and by letting the pseudo-distributions of two inliers fail with a probability near the admissible threshold (those with a failure probability greater than zero but less than $\alpha_{e}=2/49$). Note that the expanded success domain at the top-right subplot contains these inliers, whose requirement and mean values correspond to the red symbols in the bottom-left subplot. The risk-agnostic nature of the design formulation ignores these $r$ values. The bottom-right subplot shows the extent by which the elimination of two aleatoric outliers shifts the CDF to the left, thereby reducing the empirical mean, i.e., the blue-dashed vertical line is to the left of the magenta-dashed vertical line. 

\section{Robustness Analysis}\label{ua}
The formulations above synthesize a design $\theta^\star$ using computationally tractable heuristics for $\mathbb{P}_a[{\mathcal F}_k(\theta^\star,E)]\leq \alpha_{a,k}$ for $k=1,\ldots n_r$ based on the datasets ${\mathcal A}$ and ${\mathcal E}$. However, without further analyses, we cannot know if $\theta^\star$ meets the desired robustness specifications for additional scenarios. The extent by which these heuristics address the original problem can be evaluated using \emph{Robust Monte Carlo} (RMC) analysis and \emph{Robust Scenario Theory} (RST). The term ``robust'' is used to distinguish cases in which epistemic uncertainty is present. A RMC analysis estimates the range of failure probabilities corresponding to all values of $e$ in $E$ for any design $\theta$ regardless of the means by which such a design was obtained. However, the additional data required to compute such an estimate might not be available, e.g., the scenarios are obtained experimentally. On the other hand, RST yields a rigorous, distribution-free upper bound to this range for scenario-based optimal designs without requiring additional data. These methods will be introduced next.

\subsection{Robust Monte Carlo Analysis} \label{mcfea}
The desired failure probability range can be readily approximated using sampling. This process first entails generating the testing datasets ${\mathcal A}'$ and ${\mathcal E}'$ having $n_a' \gg 1$ and $n_e' \gg 1$ scenarios respectively. Whereas the scenarios in ${\mathcal A}'$ are drawn from 
$\mathbb{P}_a$, those in ${\mathcal E}'$ are drawn uniformly from $E$. Details of a nested sampling procedure are presented next. This procedure, just as the scenario programs above, enables the analyst to eliminate aleatoric and epistemic outliers from the analysis (if any). The values of $\alpha_{a,k}$ and $\alpha_{e,k}$ used to obtain $\theta^\star$ might differ from those used subsequently. To make this distinction clear we will use $\alpha'_{a,k}$ and $\alpha'_{e,k}$ instead.

Consider the sequence
\begin{equation}
{\mathcal R}_k(e^{(j)})=\left \{ r_k(\theta, a^{(i)},e^{(j)}) \right \}_{i=1}^{n'_a},
\end{equation}
for any  $e^{(j)}\in\mathcal{E}'$ and any\footnote{The total failure probability can be studied by replacing $r_k$ with $r_{\text{max}}$ in (\ref{wcr}) hereafter.} $k\in\{1,\ldots n_r\}$, and as ${\mathcal R}_k'$ the subsequence of ${\mathcal R}_k$ containing the smallest $\lceil n'_a (1-\alpha'_{a,k})\rceil$ values. The failure probability corresponding to $e$ is $p_k(e)=\mathbb{P}_a[{\mathcal F}_k(\theta,e)] \approx 1-F_{{\mathcal R}_k'(e)}(0)$. This yields the sequence 
\begin{equation}
{\mathcal P}_k=\left\{ p_k(e^{(j)})\right \}_{j=1}^{n'_e},\label{seqp}
\end{equation}
and the range \begin{equation}
a_k(\theta,{\mathcal A}', \mathcal{E}',\alpha'_{a,k},\alpha'_{e,k}):= \left [F^{-1}_{{\mathcal P}_k}(0),\, F^{-1}_{{\mathcal P}_k}(1-\alpha'_{e,k})\right ]. \label{pfrange}
\end{equation}

Hereafter, the lower and upper limits of intervals will be denoted with underlines and overlines respectively. Furthermore, the $100\sigma\%$ CDF confidence interval based the observations of $d$ in ${\mathcal D}$ evaluated at $d'$ will be denoted as CI$_{\sigma}({\mathcal D},\, d')$. 
The non-parametric framework in \cite{Hanson10} will be used to evaluate these intervals. Confidence intervals based on the $n_a'$ aleatoric scenarios yield the failure probability range 
\begin{equation}
b_k(\theta,{\mathcal A}', \mathcal{E}',\alpha'_{a,k},\alpha'_{e,k},\sigma):=[1-\max_{j=1,\ldots n'_e} \overline{\text{CI}}_{\sigma}({\mathcal R}'_k(e^{(j)}),0),\; F^{-1}_{{\mathcal D}_k}(1-\alpha'_{e,k})], \label{bb}
\end{equation}
where ${\mathcal D}_k=\{1- \underline{\text{CI}}_{\sigma}({\mathcal R}'_k(e^{(j)}),0)\}_{j=1}^{n'_e}$. Note that $\alpha'_{e,k}=0$ yields $\overline{b}_k=1-\min_j \underline{\text{CI}}_{\sigma}({\mathcal R}'_k(e^{(j)}),0)$. The width of $b_k$, which accounts for the sampling error caused by using a finite $n'_a$, converges to a fixed value as $n_a' \rightarrow \infty$. 

A means to account for the sampling error caused by using a finite number of epistemic scenarios is presented next. Our goal is to estimate the probability of violating the robustness specification $\mathbb{P}_a[{\mathcal F}_k(\theta,e)]\leq P^{\text{max}}_k$ when $e$ is uniformly distributed in a subset of  $E$ excluding ${\mathcal O}_e$, i.e.,  bounding $\mathbb{P}_e[\mathbb{P}_a[{\mathcal F}_k(\theta,e)]> P^{\text{max}}_k]$. To start off, compute the failure probability sequence 
\begin{equation}
\mathcal{Q}_k=\left \{ 1-\underline{\text{CI}}_{\sigma}\left( {\mathcal R}'_k(e^{(j)}),0)\right) \right \}_{j=1}^{n'_e}.\label{cha}
\end{equation}
Denote as ${\mathcal Q}'_k$ the subsequence of $\mathcal{Q}_k$ containing the smallest $\lfloor n'_e (1-\alpha'_{e,k})\rfloor$ elements. An estimate of the probability of not meeting the robustness specification for the epistemic scenarios prescribing ${\mathcal Q}'_k$ is
\begin{align}
\mathbb{P}_e\left [1-\underline{\text{CI}}_{\sigma}\left( {\mathcal R}'_k(e),0)\right)>P^{\text{max}}_k\right]&\approx c_k:= 1-F_{{\mathcal Q}'_k}(P^{\text{max}}_k). \label{qq}
\end{align}
This estimate falls onto the $100\sigma\%$ confidence interval 
\begin{equation}
d_k(\theta,{\mathcal A}', \mathcal{E}',\alpha'_{a,k},\alpha'_{e,k},\sigma, P^{\text{max}}):={\text{CI}}_{\sigma}\left(\mathcal{Q}'_k, \,P^{\text{max}}_k\right). \label{mixed_ci}
\end{equation}
The width of this interval converges to zero as $n_a' \rightarrow \infty$ and $n_e' \rightarrow \infty$. Note that $\overline{b}_k$, $c_k$ and $\overline{d}_k$ account for the aleatoric sampling error, but only $\overline{d}_k$ accounts for the epistemic sampling error.

Robustness specifications can be cast in terms of $\overline{a}_k$, $\overline{b}_k$, $c_k$ or $\overline{d}_k$. A given design $\theta$ might be unacceptable for $\alpha'_{e,k}=0$, but acceptable for $0<\alpha'_{e,k}\ll1$, i.e., when WC elements of $E$ are removed from the RMC analysis. Assuming $\alpha'_{e,k}>0$ might be justified by acknowledging that the prescription of $E$ is intrinsically uncertain, and only one of its elements will determine the actual system's performance.

\subsection{Robust Scenario Analysis} \label{scefea}
The developments in \cite{Garatti22} are applicable to data-driven designs subject to aleatory uncertainty only.  The following developments extend the framework in \cite{Campi25} to decision making problems subject to aleatory and epistemic uncertainty. Proofs will be presented elsewhere due to space limitations.

For simplicity in the presentation we will focus on the single requirement case after grouping both uncertainty types into $\delta=(a,e)\in {\mathbb R}^{m_a+m_e}$. This allows rewriting program (\ref{sta2}) as
\begin{align}
	 \min_{\theta \in \Theta} & \quad J\left(\theta \right) \label{sta3}\\
	\text{ subject to:} & \quad r(\theta,\delta^{(i)})\leq 0\; \text{ for all } \delta^{(i)}\in \tilde{{\mathcal D}},\nonumber 
\end{align}
where $\tilde{{\mathcal D}}\subseteq {\mathcal D}=\{\delta^{(i)}\}_{i=1}^{n_a}$. In this setting, the scenario $\delta^{(i)}$ is given by the multi-point representation $\delta^{(i)}=a^{(i)}\times\tilde{{\mathcal E}}=\{\delta^{(i,j)}\}_{j=1}^{n_e}$, where $\tilde{{\mathcal E}}\subseteq {\mathcal E}$. For instance, the aleatoric dataset $\tilde{{\mathcal A}}=\{a^{(i)}\}_{i=1}^{n'_a}$ and the epistemic dataset $\tilde{{\mathcal E}}=\{e^{(1)}, e^{(2)}, e^{(3)}\}$ yield $\delta^{(i)}=\{(a^{(i)},e^{(1)}), (a^{(i)},e^{(2)}),(a^{(i)},e^{(3)}) \}$ for $i=1,\ldots n'_a$. 
Denote as $\theta^\star({\mathcal D})$ a design resulting from any of the above formulations.\footnote{The robustness analysis of $\theta^\star$ that follows is with respect to a given epistemic set. Choose this set to be $E$ when $\alpha_{e,k}=0$ for all $k=1,\ldots n_r$. Otherwise, this set should be a large subset of $E$ excluding the epistemic outliers. This practice, which ensures that $\theta^\star$ has a high chance of satisfying the requirements for a comparatively large set of epistemic uncertainties, will yield the most informative analysis. Note, however, that the epistemic set used for design is fully decoupled from the set to be used for analysis, so any choice is acceptable.} Recall that such formulations choose  $\tilde{{\mathcal A}}$ from ${\mathcal A}$ and $\tilde{{\mathcal E}}$ from ${\mathcal E}$ according to the values for $\alpha_a$ and  $\alpha_e$ chosen.

\begin{definition}[Support scenarios]
	\label{def:ss}
	The support set of a scenario program is any subset ${\mathcal D}' \subseteq {\mathcal D}$ satisfying $\theta^\star({\mathcal D})=\theta^\star({\mathcal D}')$.  
\end{definition}

A scenario program might admit multiple support sets, with the one having the smallest cardinality being preferable. When the scenario program is convex, the number of support scenarios is less than or equal to $m_{\theta}$, and all such scenarios correspond to active constraints. 
 
\begin{definition}[Point violation and risk]
	\label{def:N}
A design $\theta$ violates the requirement at the uncertain point $\delta=(a,e)$ if $a\in \mathcal{F}(\theta,e)$. The point-risk of $\theta$, denoted as $\textnormal{Risk}_e(\theta)$, is defined as $\textnormal{Risk}_e(\theta):=\mathbb{P}_a [a\in \mathcal{F}(\theta,e)]$. 
\end{definition}

Note that $\textnormal{Risk}_e(\theta)$ is the failure probability for a fixed epistemic point $e$.  In \cite{Campi21ml}, $\textnormal{Risk}_e(\theta^\star)$ is characterized by tight bounds that can be computed from a statistic (called \emph{complexity}) of the training set ${\cal D}$, and the construction of the bounds does not require any knowledge of $\mathbb{P}_a$. Therefore, all the available data is not only used to train design $\theta^\star$ but also to provide a tight evaluation of the ensuing risk (as opposed to approaches that perform these tasks with different datasets).

\begin{definition}[Set violation and risk]
	\label{def:P}
	A design $\theta$ violates the requirement at the uncertain set $\Delta=a\times E$ if $a \in {\mathcal F}(\theta,E)$.  The set-risk of $\theta$, denoted as $\textnormal{Risk}_E(\theta)$, is defined as $\textnormal{Risk}_E(\theta):= \mathbb{P}_a \left [\, a \in {\mathcal F}(\theta,E) \right]$. 
\end{definition}

The distribution $\mathbb{P}_a$ fully prescribes the randomness of $a$, thus of $\Delta$. As we shall see, the set-risk can be upper bounded with precision by only using information available in the training set. 
Designs can be made robust by selecting a suitable multi-point sequence ${\mathcal E}$, e.g., some elements of $\tilde{{\mathcal E}}$ might fall outside $E$. Even though this choice impacts the resulting $\theta^\star$, the results hold true for any choice of $E$. 
  
A rigorous upper bound of $\textnormal{Risk}_E(\theta^\star)$ is presented below. The possibility of such an accurate evaluation is notable considering that the concept of set-risk involves the infinitely many values of $e$ in $E$, while the training set ${\mathcal E}$ only includes a finite number of them. Key to this achievement is the determination of a suitable statistic of the training set called the ``set-complexity''.  This statistic, which can be computed from ${\mathcal D}$ and $E$, is instrumental in bounding the set-risk. 

\begin{definition}[Set-complexity]
	\label{def:adv_complex}
The set-complexity of $\theta^\star({\mathcal D})$, denoted as $s_E$, is the number of elements in $\mathcal{D}$ satisfying at least one of the following conditions: \textnormal{(i)} $a^{(i)}\in {\mathcal F}(\theta^\star, E)$, and \textnormal{(ii)} $\delta^{(i)}$ is a support scenario.  
\end{definition}
The scenarios satisfying (i) do not satisfy the requirements robustly, i.e., there exists an epistemic realization $e\in E$ for which $r(\theta^\star,a^{(i)},e)>0$. Thus, (i) evaluates the empirical set-risk. Conversely, condition (ii) accounts for the training scenarios overfitted by $\theta^\star$. Hence, the set-complexity is the cardinality of the union of the sub-sample of ${\mathcal D}$ that fail to satisfy the requirement for all elements of $E$, and the support sub-sample of $\theta^\star$.

Condition (i), which entails considering the infinitely many epistemic points in $E$, cannot be evaluated exactly in many practical applications, e.g., when $r$ is only know implicitly. As such, we must resort to sampling- or optimization-based approximations, such as the one detailed in the 
section below. 
On the other hand, condition (ii) can be evaluated by solving an auxiliary scenario program: the scenario $\delta^{(i)}$ is a support scenario if $\theta^\star({\mathcal D}\setminus \delta^{(i)}) \neq \theta^\star({\mathcal D})$.

The last step in preparation of the main theorem is defining the function $\overline{\epsilon}(k)$ having domain $k \in \{0,1,\ldots n_a\}$ and codomain $[0,1]$. This function, derived in \cite{Campi21ml}, evaluated at the set-complexity yields an upper bound of the set-risk. Evaluating the bound $\overline{\epsilon}(k)$ requires choosing a confidence level $\beta\in(0,1)$, commonly assumed to be small.

\begin{definition}[Risk bound] For a given value of $\beta$ in $(0,1)$ and any $k$ in $\{0,1,\ldots,n_a-1\}$, consider the polynomial equation in the variable $t$ 
\begin{equation}
		\label{pol_eq-for-eps(k)-relax}
		{n_a \choose k}t^{n_a-k} - \frac{\beta}{2n_a} \sum_{i=k}^{n_a-1} {i \choose k}t^{i-k} - \frac{\beta}{6n_a} \sum_{i=n_a+1}^{4n_a} {i \choose k}t^{i-k}  = 0.
\end{equation}
Equation (\ref{pol_eq-for-eps(k)-relax}) has two solutions in $[0,+\infty)$, which we denote as $\underline{t}(k)$ and $\overline{t}(k)$ such that $\underline{t}(k) \leq \overline{t}(k)$. When $k=n_a$ assume $\underline{t}(n_a) = 0$. We define the desired bound as $\overline{\epsilon}(k) := 1 - \underline{t}(k)$. \hfill$\star$
\end{definition}
The zeros of (\ref{pol_eq-for-eps(k)-relax}) can be efficiently computed using the numerical approach given in \cite{Garatti22}. We are now ready to state the main result.
\begin{theorem}
	\label{main_th_standard}
	Assume that ${\mathcal D}$ is comprised of Independent and Identically Distributed (IID) samples drawn from a non-accumulating distribution ${\mathbb P}_a$. We have that 
	\begin{equation} 
		\label{eq:concentration-VS-relax}
		{\mathbb P}_a^{n_a} \left[\, \textnormal{Risk}_E(\theta^\star({\mathcal D}))\leq \overline{\epsilon}(s_E) \,\right]\,\geq\,1-\beta,
	\end{equation}
	where $\theta^\star$ is the optimal design and $s_E$ is the set-complexity. 
\end{theorem}
Hence, for any training sequence ${\mathcal D}$ using the multi-point representation ${\mathcal E}$ of $E$, the {\em set-risk} of $\theta^\star({\mathcal D})$ is bounded from above by $\overline{\epsilon}(s_E)$ with confidence $1-\beta$. The confidence refers to the possibility of collecting an unlucky training set for which the bound does not hold.

This bound is \emph{distribution-free} since it holds for any ${\mathbb P}_a$. Equation \eqref{eq:concentration-VS-relax} contains ${\mathbb P}_a$ twice, once explicitly in ${\mathbb P}_a^{n_a}$ and another one implicitly through the definition of $\textnormal{Risk}_E(\theta)$. However, this theorem can be applied without knowing this probability: for any aleatory sequence ${\mathcal A}$ and any epistemic set $E$, we first generate the multi-point sequences ${\mathcal E}$ and ${\mathcal D}$. We then compute $\theta^\star({\mathcal D})$ using one of the scenario programs above. Then, the set-complexity $s_E$ is evaluated from $\theta^\star$, ${\mathcal D}$ and $E$, from which the desired bound $\overline{\epsilon}(s_E)$ can be readily computed. The scenario bounds are particularly tight when the program is fully supported, i.e., the number of support constraints is $m_\theta$ for all ${\mathcal D}$ with $n_a\geq m_\theta$. 

Note that the number of support scenarios of the moment-based programs in Section \ref{MOM} is equal to $n_a$, thereby making $\overline{\epsilon}(s_E)=1$. As such, the scenario analysis is only suited to the moment-independent programs of Section \ref{RCC}.

\subsubsection{Set containment condition}
Sampling- and optimization-based tests for evaluating the set containment condition $a\in {\mathcal F}(\theta^\star, E)$ in Definition 4 are presented next. 

In a sampling-based test we first generate a set ${\mathcal E'}$ of $n'_e\gg1$ epistemic points uniformly distributed in $E$. A sufficient condition for containment is $r_\text{max}(\theta^\star,a,e^{(j)})> 0$ for some $e^{(j)}$ in ${\mathcal E'}$. When this is not the case, statistical arguments can be used 
(Section \ref{mcfea}), i.e., $a\in {\mathcal F}(\theta^\star, E)$ is almost always true when the upper limit of the confidence interval of the failure probability is acceptably small. 

An optimization-based test for containment is presented next. Write the epistemic set as $E=\{e: \|c-e\|\leq \nu\}$, where  $c\in \mathbb{R}^{m_e}$ is the center, $\nu\in\mathbb{R}$ is a non-negative number, and $\|\cdot\|$ is a suitable norm \cite{Crespo08}. These norms can be used to characterize various geometries including hyper-rectangles and hyper-ellipsoids. 
Consider the optimization program
\begin{align}
	\min_{e\in E} & \quad  \|c-e\| \text{ such that } r(\theta^\star,a,e)>0,\label{maxset}
\end{align}
where $ r(\theta,a,c)<0$ is assumed. Denote as $e^\star$ the solution of (\ref{maxset}). The \emph{Maximal Set}, defined as $\tilde{E}:= \{e: \|c-e\|\leq \|c-e^\star\|\}$, satisfies $a\in {\mathcal S}(\theta^\star,\tilde{E})$. Therefore, $a\in{\mathcal S}(\theta^\star,E)$ iff $\|c-e^\star\|\geq\nu$. Hence, the containment condtion $a\in {\mathcal F}(\theta^\star, E)$ holds iff $\|c-e^\star\|<\nu$. This statement is exact when the solution to (\ref{maxset}) converges to the global minimum.  

A significant advantage of the sampling test over the optimization test is that its computational demands are independent of $m_\theta$. However, the information produced by the optimization test can be used to evaluate set containment for sets epistemic sets having the same geometry of $E$ but different sizes.  

\section{Sequential Design}\label{sd}
Using large training datasets to find $\theta^\star$ increases the chance of such a design satisfying tight robustness specifications. However, this practice might increase the computational cost of finding $\theta^\star$ beyond acceptable limits. 

A design strategy amenable to computationally expensive models is outlined next. This strategy, which combines low-fidelity training with high-fidelity testing, is called \emph{Sequential Design} (SD). Each design point in this sequence is trained using small datasets ${\mathcal A}$ and ${\mathcal E}$ in order to keep the cost low. However, these sets contain elements of the large testing datasets ${\mathcal A}'$ and ${\mathcal E}'$ seeking to expand the success domain of the preceding design (see details below). 

The algorithmic implementation presented next uses (\ref{chancon_scex}) but any other of the above programs can be used instead. To start off, set the maximum number of iterations $\overline{\ell}$, the testing sets ${\mathcal A}'$ and ${\mathcal E}'$, the robustness specifications $\phi_k(\theta)$, the performance bound $\overline{J}$, the number of aleatoric scenarios violating the $k$-th requirement\footnote{ $b_k=\lceil (n_a/n_a') \sum_{i=1}^{n_a'}I[\max_{j=1,\ldots n_e'} r_k(\theta_\ell,a^{(i)},e^{(j)})>0]\rceil$ is a good choice.}, $b_k\geq 0$, and the joint density\footnote{When $\mathbb{P}_a$ is unknown, we can either learn  $f_a(a)$ from the data or assume it is constant.} of $\mathbb{P}_a$, $f_a(a)$. Furthermore, make $1 \leftarrow \ell$ and let $\theta_\ell$ be a baseline design. 
\begin{enumerate}  [label*=\arabic*.]
\item Evaluate $\phi_k(\theta_\ell)$ for $k=1,\ldots n_r$ by performing a RMC analysis using ${\mathcal A}'$ and ${\mathcal E}'$. Denote as $K\subseteq\{1,\ldots n_r\}$ the requirements for which the specifications are violated.  
\item If $(K=\emptyset\, \cap\, J(\theta_\ell)\leq \overline{J})$ or $\ell=\overline{\ell}$ then stop. 
\item If $K\neq \varnothing$ increase $n_a$ and make $\alpha_{a,k}=0$ for all $k=1,\ldots n_r$. Otherwise, increase $\alpha_{a,k}$ for all $k\in K$.
\item  Compute  ${\mathcal A}=\{a^{(i)}\in {\mathcal A}':  s^\star_i=1\}$, where $s^\star\in \mathbb{R}^{n_a'}$ is the solution to\footnote{Infeasibility is prevented by subtracting the additional decision variable $\beta_k\geq0$ to $b_k$, and adding $\rho \sum \beta_k$ with $\rho\gg1$ to the objective function.}
\begin{align}
	 \max_{s \in [0,1]} & \quad \sum_{i=1}^{n_a'} s_i \gamma_i f_a(a^{(i)}) + \lambda \det(\text{cov}({\mathcal A}',s)) \label{tra}\\
	\text{ subject to:} & \quad \sum_{i=1}^{n_a'} s_i = n_a, \quad \sum_{i=1}^{n_a'} c_{i,k} s_i = b_k\; \text{for}\; k=1,\ldots n_r,\quad s^\top (1-s)=0, \nonumber
\end{align}
$c_{i,k}=I[a^{(i)}\in {\mathcal F}_k(\theta_\ell,{\mathcal E}')]$, $\gamma_i=\max_k c_{i,k}$, $\det(\cdot)$ is the determinant and  $\text{cov}(\cdot,w)$ is the weighted covariance with weights $w\in \mathbb{R}^{n_a'}$ in the principal component axes. 
\item Find $n_e$ elements of ${\mathcal E}'$ for which $r_{\text{max}}(\theta_\ell,a^{(i)},e^{(j)})$ takes on the greatest value for all $a^{(i)}\in {\mathcal A}$. These scenarios constitute ${\mathcal E}$.
\item Solve for $\theta^\star({\mathcal A},{\mathcal E})$ in\footnote{The analyst might consider using a data-weighted extension of the inverse CDF in the Appendix, where the weight of an aleatoric scenario is its normalized likelihood.} (\ref{chancon_scex}). 
\item  Make $\ell+1 \leftarrow \ell$, $\theta_\ell \leftarrow  \theta^\star$, and go to Step 1.
\end{enumerate}

Note that the testing scenarios used for training attain $s^\star_i=1$. These scenarios attain both a high likelihood and a large separation among themselves. The balance between these two criteria can be adjusted by changing $\lambda$. By choosing failure points with high-likelihood we seek to expand the success domain where the failure probability will drop the most. The large separation property ensures that non-clustered scenarios from several potentially different failure modes are included in ${\mathcal A}$. 

By training a new $\theta^\star$ with the sequentially updated datasets we seek to expand the success domain of the preceding design. Note however that the underlying dependency of the requirements on $\theta$, $a$, and $e$ might render tight robustness specifications infeasible regardless of the number of training scenarios.  Further notice that the training set ${\mathcal A}$ used by the SD algorithm is not comprised of IID scenarios, thereby invalidating the scenario bound.

\begin{figure*}[h!]
    \centering
    \begin{subfigure}[b]{0.325\textwidth}
        \centerline{\includegraphics[trim={21cm 2.25cm 1.5cm 2.5cm},clip,width=1.1\textwidth]{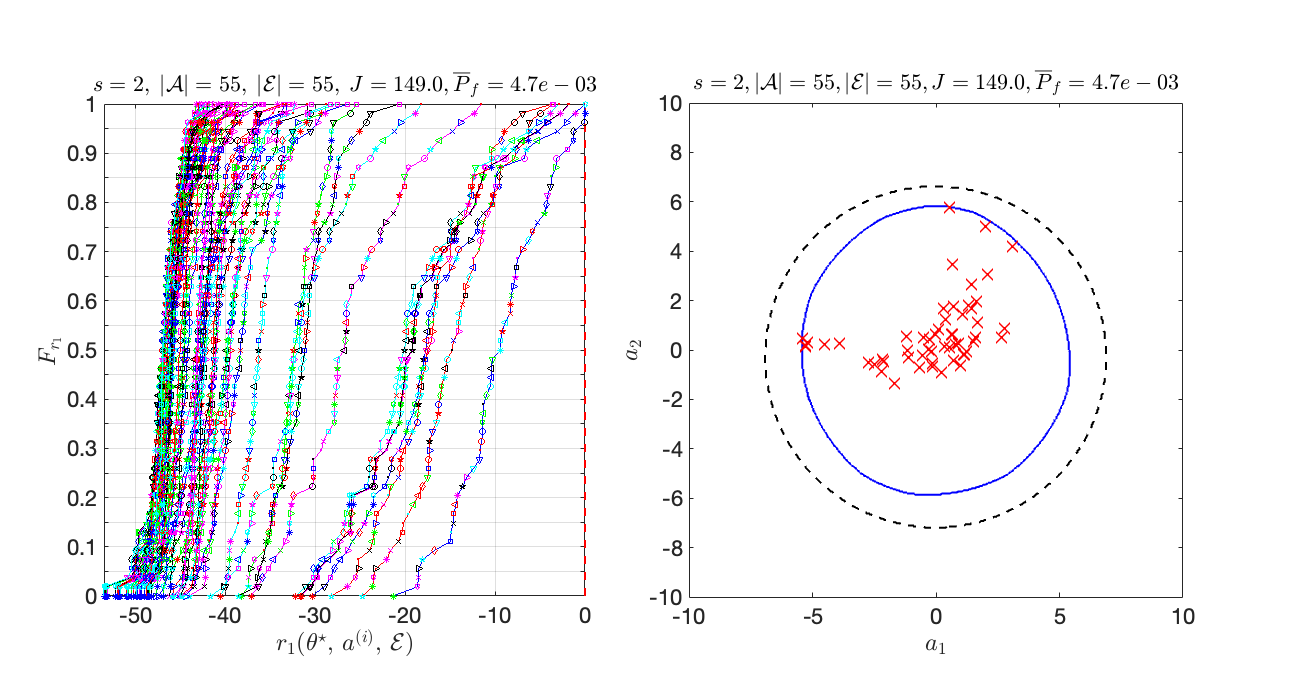}}
    \end{subfigure}
    \begin{subfigure}[b]{0.325\textwidth}
        \centerline{\includegraphics[trim={21cm 2.25cm 1.5cm 2.5cm},clip,width=1.1\textwidth]{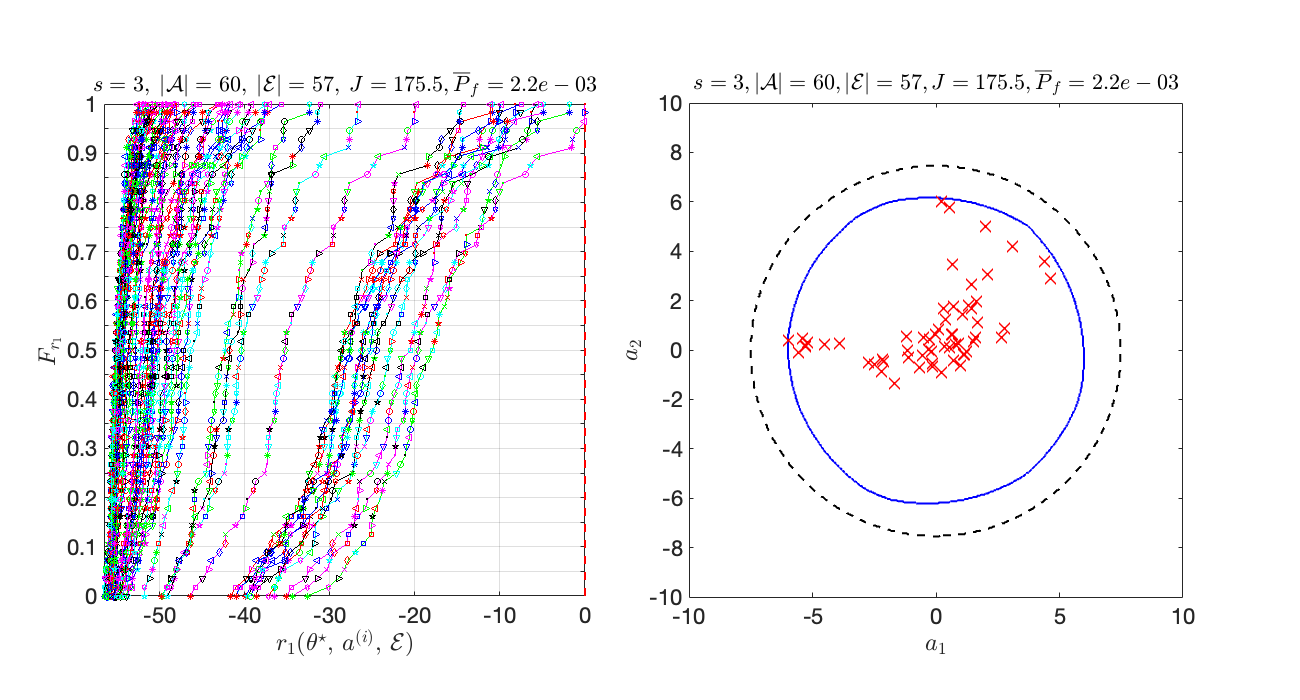}}
    \end{subfigure}
    \begin{subfigure}[b]{0.325\textwidth}
        \centerline{\includegraphics[trim={21cm 2.25cm 1.5cm 2.5cm},clip,width=1.1\textwidth]{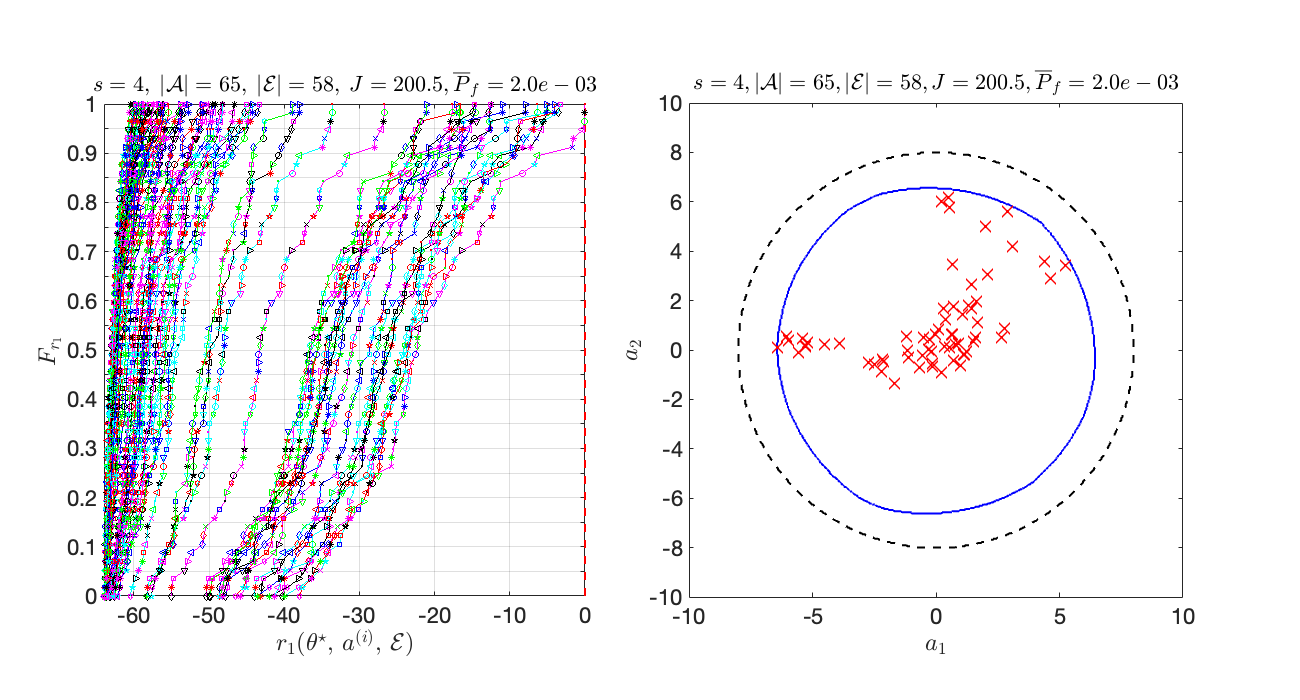}}
    \end{subfigure}
    \begin{subfigure}[b]{0.325\textwidth}
        \centerline{\includegraphics[trim={21cm 2.25cm 1.5cm 2.5cm},clip,width=1.1\textwidth]{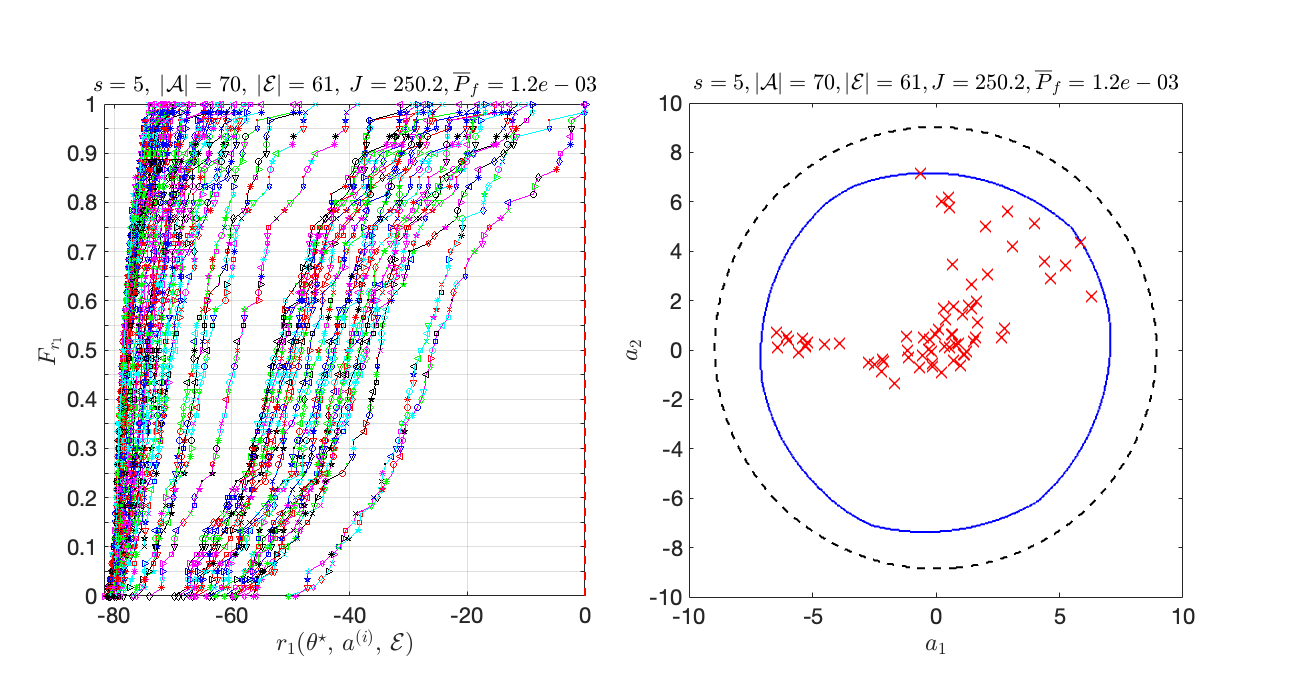}}
    \end{subfigure}
    \begin{subfigure}[b]{0.325\textwidth}
        \centerline{\includegraphics[trim={21cm 2.25cm 1.5cm 2.5cm},clip,width=1.1\textwidth]{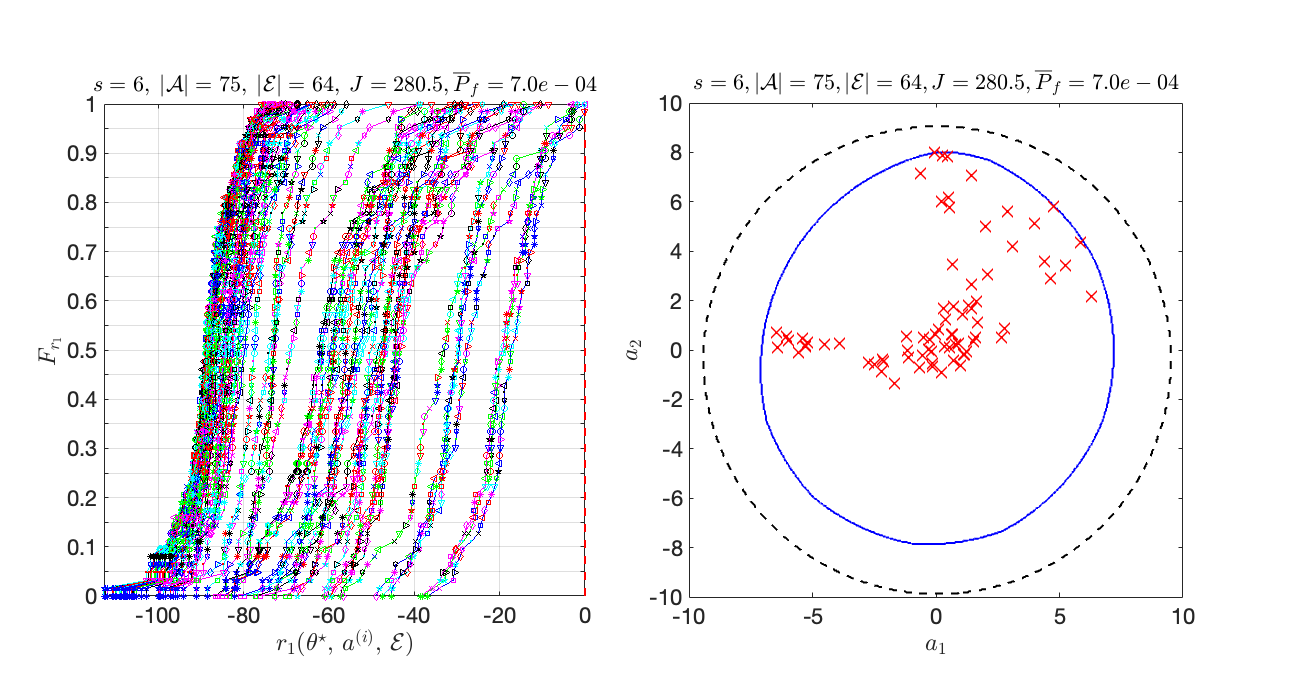}}
    \end{subfigure}
    \begin{subfigure}[b]{0.325\textwidth}
        \centerline{\includegraphics[trim={21cm 2.25cm 1.5cm 2.5cm},clip,width=1.1\textwidth]{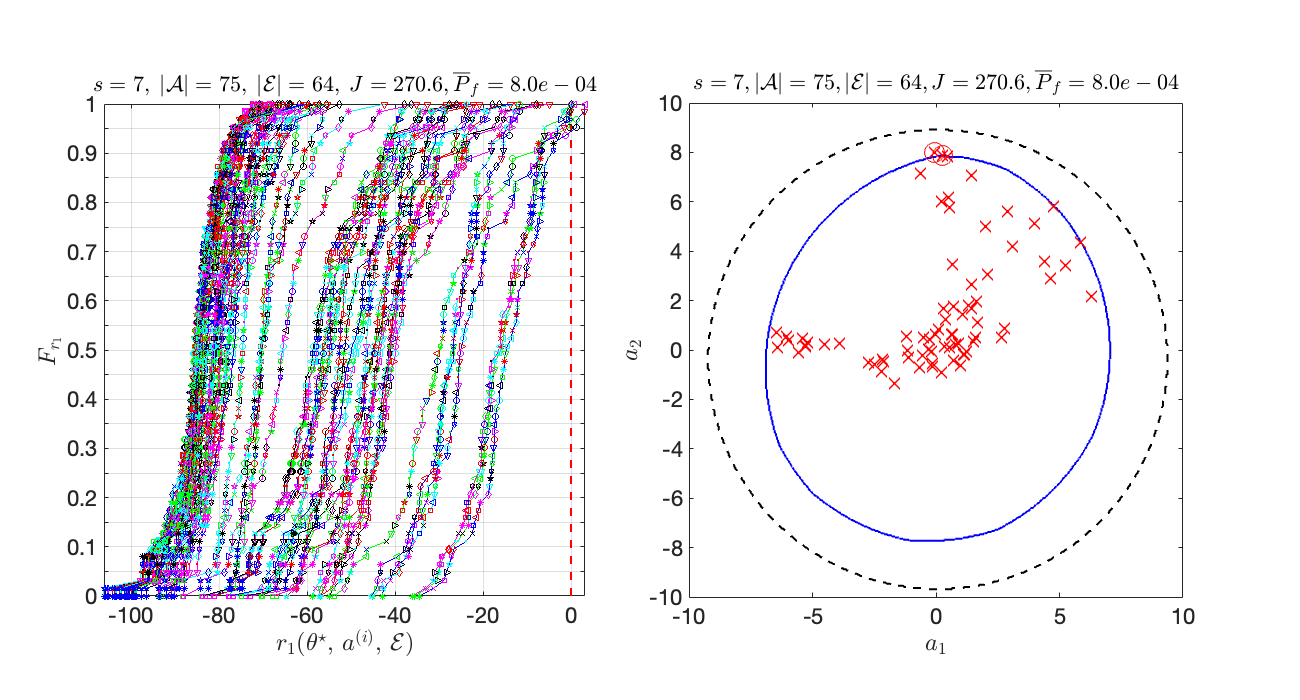}}
    \end{subfigure}
    \begin{subfigure}[b]{0.325\textwidth}
        \centerline{\includegraphics[trim={21cm 1.25cm 1.5cm 2.5cm},clip,width=1.1\textwidth]{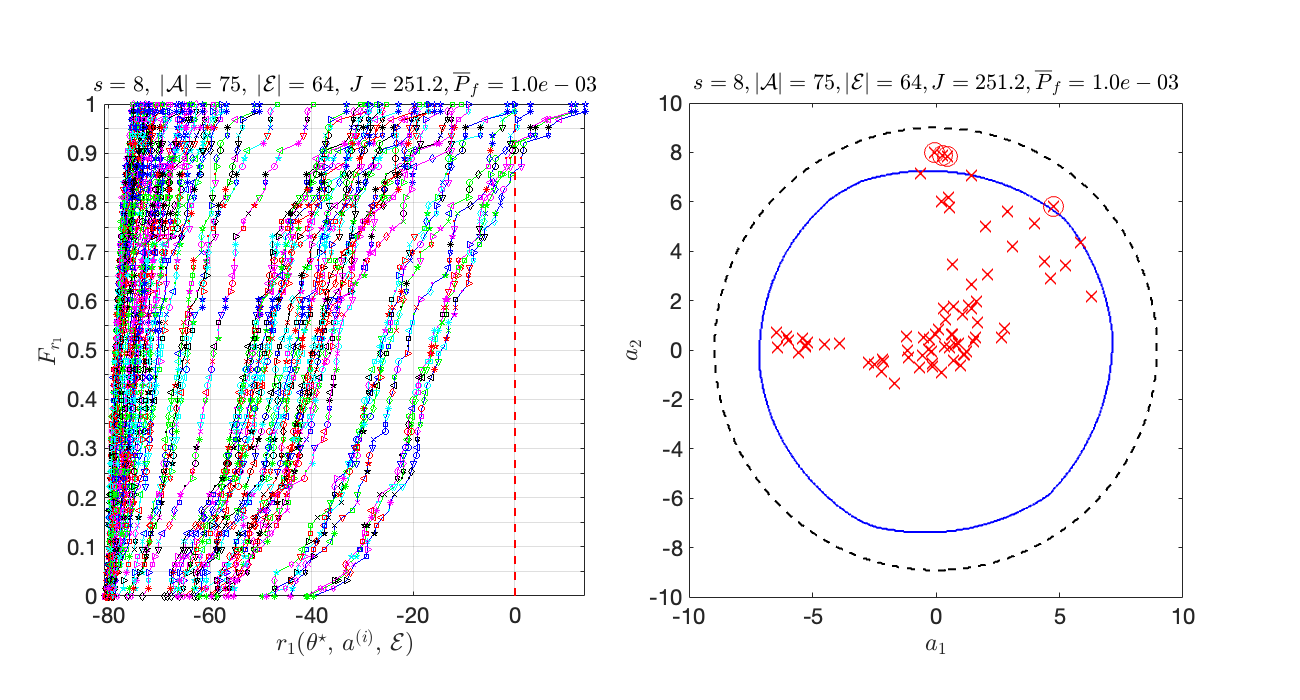}}
    \end{subfigure}
    \begin{subfigure}[b]{0.325\textwidth}
        \centerline{\includegraphics[trim={21cm 1.25cm 1.5cm 2.5cm},clip,width=1.1\textwidth]{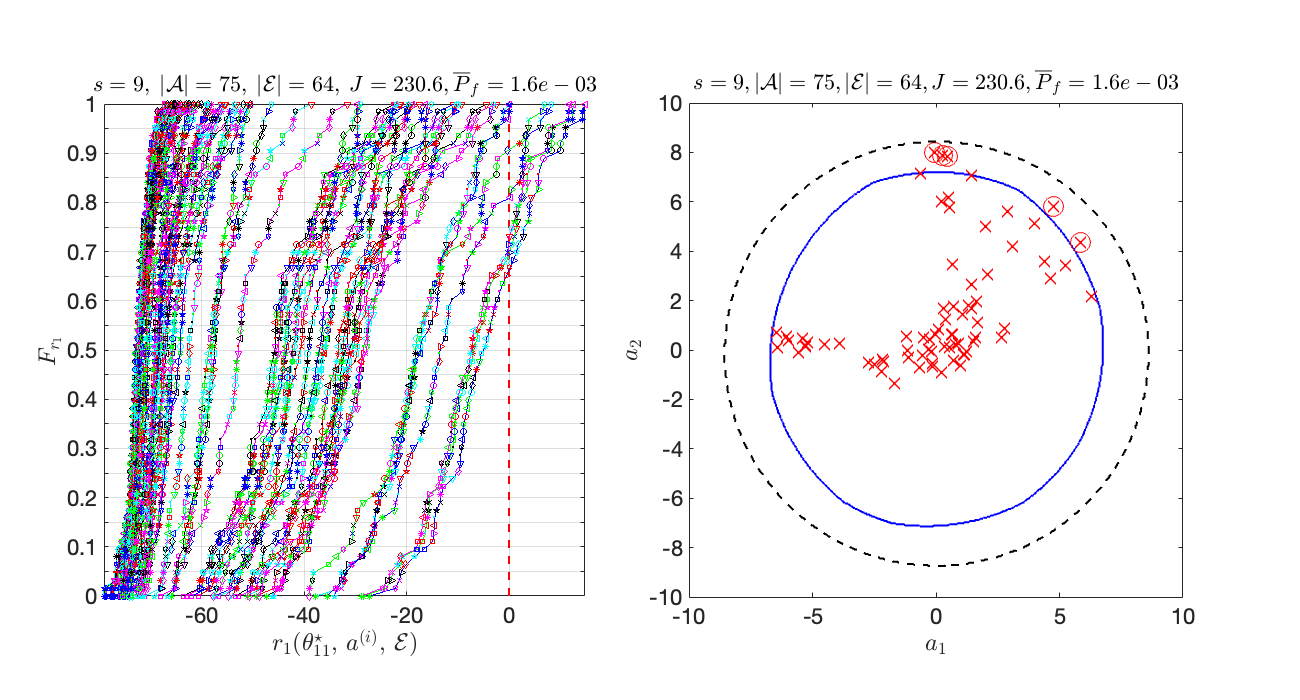}}
    \end{subfigure}
    \begin{subfigure}[b]{0.325\textwidth}
        \centerline{\includegraphics[trim={21cm 1.25cm 1.5cm 2.5cm},clip,width=1.1\textwidth]{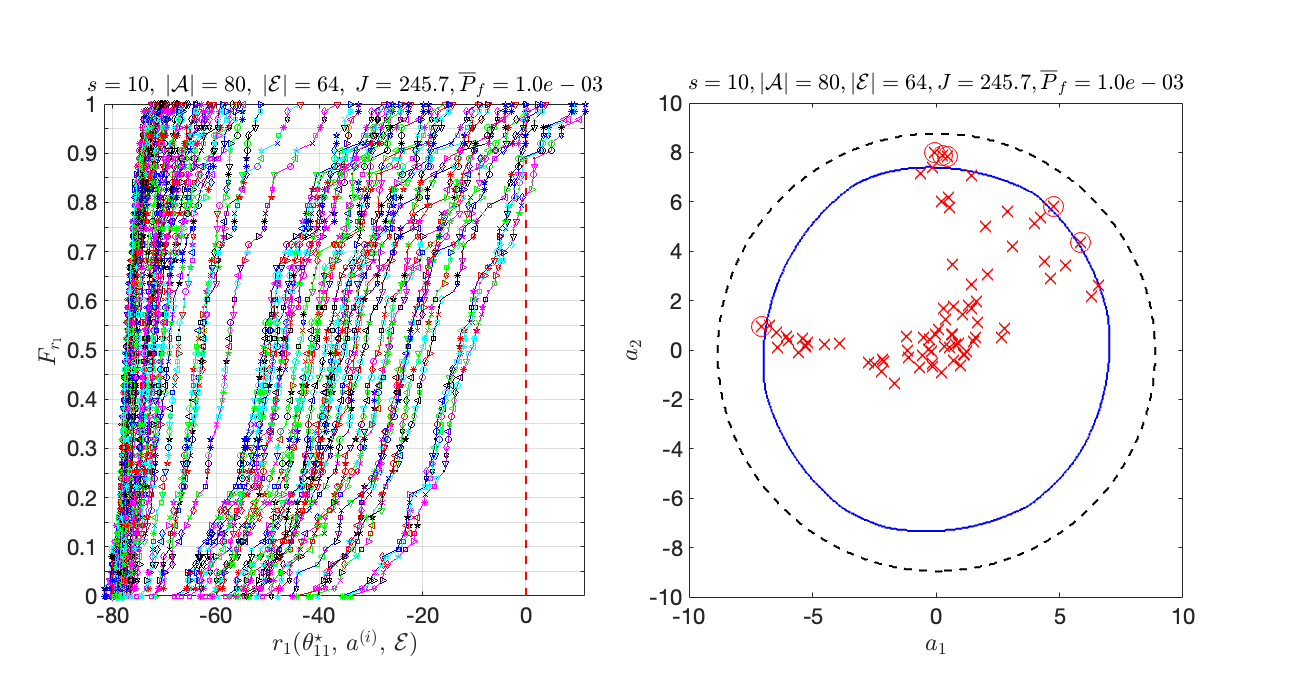}}
    \end{subfigure}
    \caption{Aleatory spaces corresponding to the design sequence resulting from the SD algorithm. Previous conventions apply. Each subplot shows the iteration number, $s$, the size of the training datasets, $\mathcal{A}$ and  $\mathcal{E}$, the objective $J$, and the failure probability $\overline{P}_f=\overline{a}(\theta_s,{\mathcal A}', \mathcal{E}',\alpha'_{a},0)$.}
    \label{fig_dalg}
\end{figure*}

\vspace{0.2cm}
\noindent {\bf Example 3 (sequential design of a data-enclosing set): } Next we impose the robustness specification $\overline{a}(\theta,{\mathcal A}', \mathcal{E}',\alpha'_{a},0)\leq P^{\text{max}}=\num{1e-3}$ to the data-enclosing set of Example 1, and assume testing sets with $n_a'=10000$ and $n_e'=200$ samples. Table \ref{tab} presents the evaluation of this specification for several designs. Note that all the designs presented thus far violate the specification. Next, we use the SD algorithm to derive $\theta_{11}^\star$ by using $\theta_1^\star$ as the baseline design and $\alpha_{e}=0$. Figure \ref{fig_dalg} shows the sequence of designs leading to $\theta^\star_{11}$, whose aleatoric space is shown at the bottom right. Note that the number of aleatoric training scenarios increases when the failure probability of the preceding design is above $P^{\text{max}}$, whereas the number outliers increases when the failure probability of the preceding design is below $P^{\text{max}}$. The elimination of outliers is instrumental to lower the objective value. More importantly, the solution to 10 scenario programs, each having no more than $n_a=80$ training scenarios rendered a design whose robustness is better than that of $\theta^\star_8$, which was trained with $n_a=5000$ scenarios. This illustrates the computational advantages of the SD algorithm. 

As expected, the resulting value for $\alpha_{a}$ is not a good estimate of the failure probability because the training scenarios are no longer IID samples drawn from $\mathbb{P}_a$. For instance, $\alpha_{a}=6/80$ was used to synthesize $\theta^\star_{11}$ in the $s=10$ iteration, but the upper limit of the failure probability range is $\overline{a}(\theta,{\mathcal A}',{\mathcal E}',\alpha_{a},0)<\num{1e-3}$.
\begin{figure*}[h!]
    \centering
    \begin{subfigure}[b]{0.25\textwidth}
        \centerline{\includegraphics[trim={21cm 1.5cm 1.1cm 3.2cm},clip,width=1.82\textwidth]{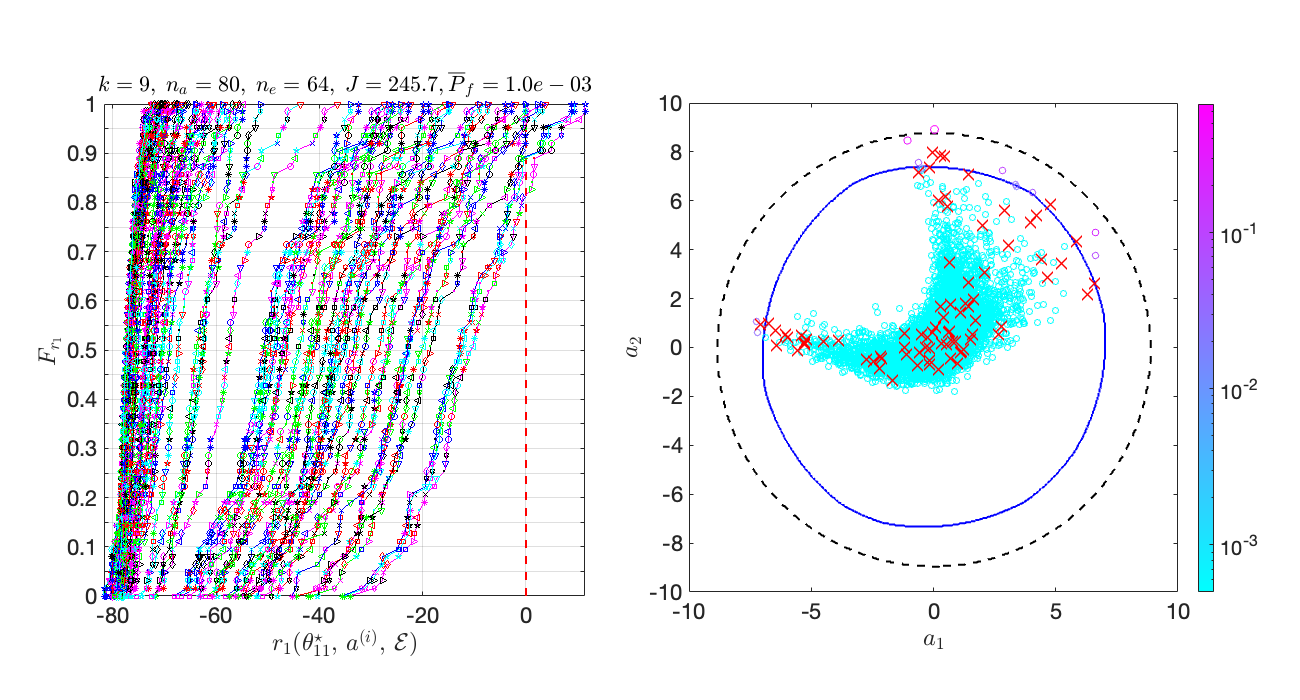}}
    \end{subfigure}
    \hspace{2cm}
    \begin{subfigure}[b]{0.25\textwidth}
        \centerline{\includegraphics[trim={23cm 1.5cm 1.1cm 2.5cm},clip,width=1.67\textwidth]{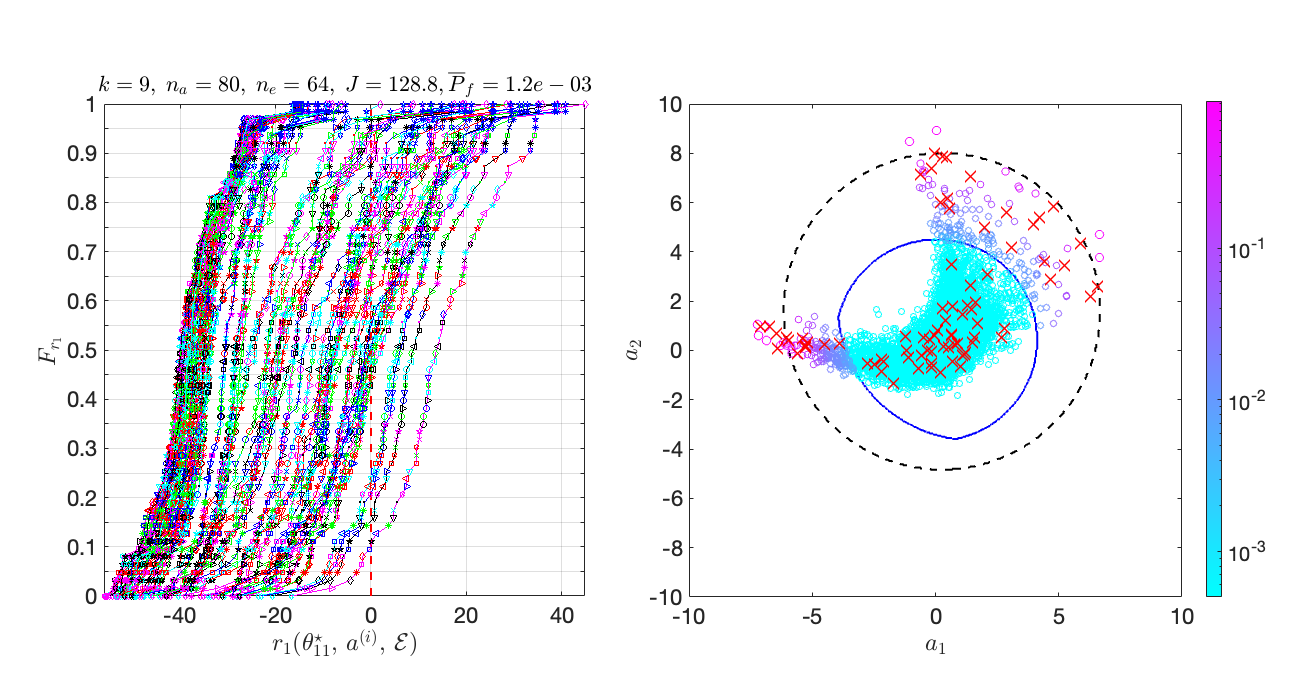}}
    \end{subfigure}
    \hfill
    \caption{Aleatory space for $\theta_{11}^\star$ (left) and $\theta_{12}^\star$ (right). Training scenarios are shown as red crosses whereas testing scenarios are colored circles. This coloring depends on the value taken by $\max_{j=1,\ldots n'_e} \mathbb{P}_e[r(\theta,a^{(i)},e^{(j)})> 0]$. The color at the bottom corresponds to zero.}
    \label{fig_dmc}
\end{figure*}

Figure \ref{fig_dmc} shows the aleatory space for design $\theta^\star_{11}$ and for the epistemically agnostic design $\theta^\star_{12}$ synthesized using ${\mathcal E}=\{(0,0,0)\}$. Each testing scenario is colored according to the fraction of the scenarios in ${\mathcal E}'$ leading to a requirement violation. As expected, the success domain of $\theta^\star_{11}$ is not only larger than the success domain of $\theta^\star_{12}$ but also encloses most of the points for which this fraction is small. Note that this fraction increases sharply with the distance between the testing aleatoric scenario and the success domain. Designs that ignore the epistemic uncertainty often fail to meet the robustness specifications regardless of how large $n_a$ is. 

\vspace{0.2cm}
\noindent {\bf Example 4 (robustness analyses): } Robustness analyses of the above designs are presented next. 

RMC analyses with $\alpha'_a=0$ and $\alpha'_e=0$ yields the values listed middle rows of Table \ref{tab}. Since many designs are based on a small number of training scenarios, their testing failure probabilities are large. In practice, the analyst should choose a design among a few alternatives according to the desired balance between performance, as measured by the objective $J(\theta^\star)$, and robustness, as measured by either the greatest failure probability, $\overline{a}(\theta,{\mathcal A}',{\mathcal E}',0,0)$, the upper limit of the corresponding aleatoric confidence interval, $\overline{b}(\theta,{\mathcal A}',{\mathcal E}',0,0,\sigma)$, or the upper limit of the mixed confidence interval $\overline{d}(\theta,{\mathcal A}', \mathcal{E}',0,0,\sigma, P^{\text{max}})$. For instance, the best design satisfying $\overline{a}<0.05$ is $\theta_{5}^\star$ since it attains the lowest objective value ($\theta^\star_{10}$ uses a different objective).  All designs violate the more comprehensive specification $\overline{d}\leq \gamma$ for $\gamma=0.01$. The $\overline{d}$ values depend on the true robustness of $\theta$ and the chosen value for $n'_a$ and $n'_e$. The greater $n'_a$ and $n'_e$, the more accurate the sampling-based estimate and the smaller $\overline{d}-\underline{d}$. Therefore, designs for which $\underline{d}\leq \gamma \leq\overline{d}$ should be evaluated by using larger testing sets. A RMC analysis with $n_a'=\num{1e4}$ and $n_e'=\num{1e3}$ samples led to $\overline{d}=\num{3.68e-3}<\gamma$ for $\theta^\star_1$, $\theta^\star_7$, $\theta^\star_8$ and $\theta^\star_{11}$. Hence, design $\theta^\star_1$, which was synthesized using small training sets, is the best choice since it attains the lowest cost among all sufficiently robust designs.

RST analyses are presented next. The corresponding results are shown at the bottom of Table \ref{tab}. Recall that this framework is inapplicable to designs using non-IID data, thereby leading to the Non-Applicable (NA) entry. The number of scenarios satisfying conditions (i) and (ii) in the definition of the set-complexity are listed as $n_s$ and $n_v$ respectively, whereas $s_E$ and $\overline{\epsilon}(s_E)$ are the set-complexity and the desired bound respectively. To this end, the set containment condition is evaluated by using the optimization-based test. Hence, an upper bound of the set-risk of $\theta^\star_8$ violating the requirement is less than \num{4.4e-3} with $1-\beta=0.9999$ confidence. Thus, increasing the size of the training dataset from $n_a=100$ (used to synthesize $\theta^\star_7$) to $n_a=5000$ (used to synthesize $\theta^\star_8$) reduced the set-risk bound 324 times. Conversely, increasing $n_e$ might not necessarily render a reduction in $\overline{\epsilon}(s_E)$. The trends seen in the failure probability are generally consistent with those seen in the bound. Whereas the failure probability estimate requires evaluating the requirement function $n'_a\times n'_e$ times, the bound requires solving $n_a$ scenario programs to find the support set, and $n_a$ optimization programs (\ref{maxset}). When the aleatory scenarios required by the Monte Carlo analysis are available and the cost of evaluating the requirements is moderate, a RMC analysis is preferable. Otherwise, the RST bound will be the only choice. Note that this bound becomes tighter as $n_a$ increases. 

Recall that RST decouples the prescription of the epistemic set $E$, thus of ${\mathcal E}$, used in training from the prescription of the epistemic set $E'$, thus of ${\mathcal E'}$ used in testing. We will use this property to evaluate the robustness of $\theta^\star_6$ with respect to the epistemic set $E\setminus \hat{E}$, where $\hat{E}$ is a small set containing the outliers. Note that the elimination of two aleatoric outliers and two local epistemic outliers for each pseudo-distribution reduced the objective 49.6\% (from 137.9 to 68.4). This improvement caused a significant loss in robustness, as measured by increases in both $\overline{d}$ and $\overline{\epsilon}$. Next, we recompute these two metrics by using $E\setminus \hat{E}$ instead. We choose a set $\hat{E}$ having $5\%$ of the volume of $E$. 
This practice leads to $\overline{a}=\num{4.52e-2}$ and $\overline{\epsilon}(s_E)=0.477$, which are 27\% and 9\% lower than those based on $E$. Therefore, a 5\% reduction in the epistemic uncertainty yielded performance and robustness improvements of 49.6\% and $27\%$ respectively. This illustrates the benefits of reducing the epistemic uncertainty.

An acceptable design might violate robustness specifications cast in terms of the metrics in Section \ref{mcfea} because $n'_a$ and $n'_e$ are insufficiently large. On the other hand, the greater the number of scenarios $n_a$ used to train $\theta^\star$, the tighter the bound $\overline{\epsilon}(s_E)$. The large discrepancies between RMC and RST results observed are a consequence of the estimates resulting from the former analysis depending on the fidelity of the testing process, whereas those for the latter analysis depending on the fidelity of the training process. 

\section{Concluding Remarks} 
This paper proposes a data-driven framework to design in the presence of aleatoric and epistemic uncertainties. This framework includes a computationally cheap design strategy in which the training datasets are sequentially updated until the desired trade-off between performance and robustness is achieved. Furthermore, we present Monte Carlo- and Scenario-based strategies to evaluate the robustness of the resulting designs. The latter approach is significant because it accounts for the infinitely many realizations the epistemic uncertainty might take (even though such a design is only trained with a finite number of them). The optimization programs proposed, some of which have a number of decision variables that do not increase with the size of the dataset, can be solved using standard gradient-based algorithms while being applicable to continuous but otherwise arbitrary requirement functions. This setting is amenable to many problems in science and engineering for which the objective function and constraints, only known implicitly, must be evaluated by numerical simulation.

Strategies that model aleatory and epistemic uncertainties as distributions in a single loop often lead to spurious analyses and designs. This practice might make the predicted failure probabilities considerably smaller than the actual ones. Note that the computed designs are optimal for the full ensemble of epistemic points in ${\mathcal E}$. However, the actual performance of the system does not depend on such points but instead on the true unknown value of the epistemic parameter $e_{\text{true}}$.  As such, there might exist a design that outperforms all other designs when $e=e_{\text{true}}$. However, such a design might not even meet the robustness specifications prescribed upfront.
   

\section*{Appendix}
\subsection*{A1. Approximation to the CDF and its inverse}
Approximations to the CDF of a random variable and its inverse are presented next. Consider the non-decreasing sequence ${\mathcal Z}=\{z_i\}_{i=1}^{n}$ that results from evaluating the function $z(a; \theta)$ at $\{a^{(i)}\}_{i=1}^n$ for a fixed $\theta$ and sorting the resulting values so $z_i=z(a^{(j)})$ for some $j$ and\footnote{The function $z({a;\theta})$ could for instance be a requirement function. When $z(a^{(i)})$ for $i=1,\ldots n$ contains repeated values, Equations (\ref{Flin}) and (\ref{Finvlin}) can be used after breaking the ties with small perturbations.} $z_i < z_{i+1}$.  A continuous, piecewise linear approximation to the CDF of $z(a)$ based on ${\mathcal Z}=\{z(a^{(i)})  \}_{i=1}^n$ is
\begin{align}\label{Flin}
F_{{\mathcal Z}(\theta)}(z):=\begin{cases}
0 & \text{if } z\leq z_1,\\
\frac{1}{n-1}\left(i-1 + \frac{z-z_i}{z_{i+1}-z_i}\right) & \text{if } z_i < z \leq z_{i+1}, \\
1 & \text{otherwise.}
\end{cases}
\end{align}
The inverse of (\ref{Flin}) is
\begin{align}\label{Finvlin}
F^{-1}_{{\mathcal Z}(\theta)}(\alpha)=\begin{cases}
z_1 & \text{if } \alpha=0,\\
z_{i}+(z_{i+1}-z_i)\left((n-1) \alpha-i+1\right) & \text{if } 0<\alpha<1, \\
z_n & \text{otherwise,}
\end{cases}
\end{align}
where $\alpha\in[0,1]$ and $i=\argmin_{j=1,\dots n} \{(n-1)\alpha-j+1: j-1\leq \alpha(n-1)\}$. The approximations (\ref{Flin}) and (\ref{Finvlin}) are differentiable in $\theta$ when $z(a; \theta)$ is $C^1$ in $\theta$. This property makes standard gradient-based algorithms applicable to the above optimization programs. Note however that these functions have derivative discontinuities at the values of $\theta$ where the order of the elements in $\{z(a^{(i)},\theta) \}_{i=1}^n$ enclosing $z$ change.

\subsection*{A2. Epistemic weights}
A rule to compute the weights ${\mathcal W}(\theta,k,\alpha_{a,k},\alpha_{e,k})=\{w_{\theta,k}^{(j)}\}_{j=1}^{n_e}$ corresponding to the $k$-th requirement function $r_k$, the design point $\theta$, the datasets ${\mathcal A}$ and ${\mathcal E}$, and the fractions $\alpha_{a,k}$ and $\alpha_{e,k}$ is detailed next. 
\begin{enumerate}
\item Calculate the set ${\mathcal I}_a=\{i\in\{1,\ldots n_a\}: p_k^{(i)}\leq F^{-1}_{\{p_k^{(i)}\}_{i=1}^{n_a}}(1-\alpha_{a,k})\}$, where $p_k^{(i)}=1-F_{\{r_k(\theta,\, a^{(i)},\,e^{(j)})\}_{j=1}^{n_e}}(0)$ is the empirical failure probability corresponding to the $i$-th pseudo-distribution. Therefore, the elements of ${\mathcal I}_a$ attain the lowest $n_a(1-\alpha_{a,k})$ failure probabilities.
\item Calculate the WC requirement value $v_k^{(j)}=\max_{i \in {\mathcal I}_a} r_k(\theta,a^{(i)}, e^{(j)})$ for all $j=1,\ldots n_e$, and the quantile $s_k=F_{\{v_k^{(j)}\}_{j=1}^{n_e}}^{-1}(1-\alpha_{e,k})$. The desired weight is given by $w_{\theta,k}^{(j)}=\exp(-\gamma \max\; ([0,v_k^{(j)}-s_k]))$, where $\gamma\geq1$ and $j=1,\ldots n_e$.
\end{enumerate}
 Hence, the weight is one when $v_k^{(j)}\leq s_k$. Otherwise, it approaches zero exponentially fast as $v_k^{(j)}-s_k$ increases.

\bibliographystyle{elsarticle-num} 
 \bibliography{nref}

\begin{thebibliography}{10}
\expandafter\ifx\csname url\endcsname\relax
  \def\url#1{\texttt{#1}}\fi
\expandafter\ifx\csname urlprefix\endcsname\relax\def\urlprefix{URL }\fi
\expandafter\ifx\csname href\endcsname\relax
  \def\href#1#2{#2} \def\path#1{#1}\fi

\bibitem{Dantzig55}
G.~Dantzig, Linear programming under uncertainty, Management Science 1~(3)
  (1955) 197--206.

\bibitem{Charnes59}
A.~Charnes, W.~Cooper, Chance-constrained programming, Management Science 6~(1)
  (1959) 73--79.

\bibitem{Elghaoui98}
L.~El-Ghaoui, F.~Oustry, H.~Lebret, Robust solutions to uncertain semidefinite
  programs, {SIAM} journal of optimization 9 (1998) 33--52.

\bibitem{Ben-Tal00}
A.~Ben-Tal, A.~Nemirovski, Robust solutions of linear programming problems
  contaminated with uncertain data, Mathematical Programming 88 (2000)
  411--424.

\bibitem{Bertsimas04}
D.~Bertsimas, D.~Pachamanova, M.~Sim, Robust linear optimization under general
  norms, Operation Research Letters 32~(6) (2004) 510--516.

\bibitem{Ben-Tal09}
A.~Ben-Tal, L.~E. Ghaoui, A.~Nemirovski, Robust Optimization, Princeton
  University Press, 2009.

\bibitem{Bertsimas11}
D.~Bertsimas, D.~Brown, C.~Caramanis, Theory and applications of robust
  optimization, {SIAM review} (2011).

\bibitem{Tsatsanis98}
M.~Tsatsanis, Z.~Xu, Performance analysis of minimum variance cdma receivers,
  {IEEE} Transactions on Signal Processing 46~(11) (1998) 3014--3022.
\newblock \href {https://doi.org/10.1109/78.726814}
  {\path{doi:10.1109/78.726814}}.

\bibitem{Mehlawat21}
M.~K. Mehlawat, P.~Gupta, A.~Z. Khan, Portfolio optimization using higher
  moments in an uncertain random environment, Information Sciences 567 (2021)
  348--374.

\bibitem{Hammond24}
J.~Hammond, L.~Crespo, F.~Montemoli, A distributionally robust data-driven
  framework to reliability analysis, Structural Safety 111~(102501) (2024).

\bibitem{Shapiro09}
A.~Shapiro, D.~Dentcheva, A.~Ruszczynski., Lectures on stochastic programming:
  modeling and theory, {SIAM}, Philadelphia, PA, 2009.

\bibitem{Coit19}
D.~Coit, E.~Zio, The evolution of system reliability optimization, Reliability
  Engineering and System Safety 192 (2019) 106259.

\bibitem{Luedtke10}
J.~Luedtke, S.~Ahmed, G.~Nemhauser, An integer programming approach for linear
  programs with probabilistic constraints, in: Mathematical Programming, Vol.
  122, 2010, pp. 247--272.

\bibitem{Qiu14}
F.~Qiu, S.~Ahmed, S.~S. Dey, L.~A. Wolsey, Covering linear programming with
  violations, {INFORMS} Journal on Computing 26~(3) (2014) 531--546.

\bibitem{Chapman22}
M.~Chapman, M.~Faub, K.~Smith, On optimizing the conditional value-at-risk of a
  maximum cost for risk-averse safety analysis, {IEEE} Transactions on
  Automatic Control (2022).

\bibitem{Kataoka63}
S.~Kataoka, A stochastic programming mode, Econometrica 31~(1) (1963).

\bibitem{Lagoa99}
C.~M. Lagoa, On the convexity of probabilistically constrained linear programs,
  in: Proceedings of the 38th {IEEE} Conference on Decision and Control (Cat.
  No.99CH36304), Vol.~1, 1999, pp. 516--52.

\bibitem{Calafiore06}
G.~Calafiore, M.~Campi, The scenario approach to robust control design, IEEE
  Transactions on automatic control 51~(1) (2006) 742--753.

\bibitem{Henrion08}
R.~Henrion, C.~Strugarek, Convexity of chance constraints with independent
  random variables, Computational Optimization and Applications 41 (2008)
  263--276.

\bibitem{Henrion11}
R.~Henrion, C.~Strugarek, Convexity of Chance Constraints with Dependent Random
  Variables: The Use of Copulae, Springer New York, New York, NY, 2011, pp.
  427--439.

\bibitem{Prekopa11}
A.~Prekopa, K.~Yoda, M.~M. Subasi, Uniform quasi-concavity in probabilistic
  constrained stochastic programming, Operations Research Letters 39~(3) (2011)
  188--192.

\bibitem{Van15}
W.~van Ackooij, Eventual convexity of chance constrained feasible sets,
  Optimization 64~(5) (2015) 1263--1284.

\bibitem{ZIO2009125}
E.~Zio, Reliability engineering: Old problems and new challenges, Reliability
  Engineering \& System Safety 94~(2) (2009) 125 -- 141.

\bibitem{COIT2019106259}
D.~W. Coit, E.~Zio, The evolution of system reliability optimization,
  Reliability Engineering \& System Safety 192 (2019) 106259.

\bibitem{ENEVOLDSEN1994169}
I.~Enevoldsen, J.~Sørensen, Reliability-based optimization in structural
  engineering, Structural Safety 15~(3) (1994) 169 -- 196.

\bibitem{8272792}
R.~{Rocchetta}, E.~{Patelli}, Stochastic analysis and reliability-cost
  optimization of distributed generators and air source heat pumps, in: 2nd
  International Conference on System Reliability and Safety (ICSRS), no. 2017,
  2017, pp. 31--35.

\bibitem{YUAN2014107}
X.~Yuan, Z.~Lu, Efficient approach for reliability-based optimization based on
  weighted importance sampling approach, Reliability Engineering \& System
  Safety 132 (2014) 107 -- 114.

\bibitem{SHAN20081218}
S.~Shan, G.~G. Wang, Reliable design space and complete single-loop
  reliability-based design optimization, Reliability Engineering \& System
  Safety 93~(8) (2008) 1218 -- 1230.

\bibitem{MENG201995}
Z.~Meng, B.~Keshtegar, Adaptive conjugate single-loop method for efficient
  reliability-based design and topology optimization, Computer Methods in
  Applied Mechanics and Engineering 344 (2019) 95 -- 119.

\bibitem{WANG2020113436}
Y.~Wang, P.~Hao, H.~Yang, B.~Wang, Q.~Gao, A confidence-based reliability
  optimization with single loop strategy and second-order reliability method,
  Computer Methods in Applied Mechanics and Engineering 372 (2020) 113436.

\bibitem{YAO201328}
W.~Yao, X.~Chen, Y.~Huang, M.~{van Tooren}, An enhanced unified uncertainty
  analysis approach based on first order reliability method with single-level
  optimization, Reliability Engineering \& System Safety 116 (2013) 28 -- 37.

\bibitem{TORII2019106499}
A.~Torii, R.~Lopez, L.~Miguel, A second order {SAP} algorithm for risk and
  reliability based design optimization, Reliability Engineering \& System
  Safety 190 (2019) 106499.

\bibitem{ChenSORA2002}
X.~Du, W.~Chen, Sequential optimization and reliability assessment method for
  efficient probabilistic design, Proceedings of the ASME Design Engineering
  Technical Conference 2 (2002) 871--880.

\bibitem{LI2010384}
H.-S. Li, S.-K. Au, Design optimization using subset simulation algorithm,
  Structural Safety 32~(6) (2010) 384 -- 392, modeling and Analysis of Rare and
  Imprecise Information.

\bibitem{DEANGELIS2015170}
M.~de~Angelis, E.~Patelli, M.~Beer, Advanced line sampling for efficient robust
  reliability analysis, Structural Safety 52 (2015) 170 -- 182.

\bibitem{CHAUDHURI2020106853}
A.~Chaudhuri, B.~Kramer, K.~E. Willcox, Information reuse for importance
  sampling in reliability-based design optimization, Reliability Engineering \&
  System Safety 201 (2020) 106853.

\bibitem{NIKOLAIDIS1988781}
E.~Nikolaidis, R.~Burdisso, Reliability based optimization: A safety index
  approach, Computers \& Structures 28~(6) (1988) 781 -- 788.

\bibitem{Kuschel1997}
N.~Kuschel, R.~Rackwitz, Two basic problems in reliability-based structural
  optimization, Mathematical Methods of Operations Research 46~(3) (1997)
  309--333.

\bibitem{Cizelj1994}
L.~Cizelj, B.~Mavko, H.~Riesch-Oppermann, Application of first and second order
  reliability methods in the safety assessment of cracked steam generator
  tubing, Nuclear Engineering and Design (1994) 359--368.

\bibitem{SCHUELLER2004463}
G.~Schuëller, H.~Pradlwarter, P.~Koutsourelakis, A critical appraisal of
  reliability estimation procedures for high dimensions, Probabilistic
  Engineering Mechanics 19~(4) (2004) 463 -- 474.

\bibitem{LI20108966}
J.~Li, D.~Xiu, Evaluation of failure probability via surrogate models, Journal
  of Computational Physics 229~(23) (2010) 8966 -- 8980.

\bibitem{CHEN2013233}
P.~Chen, A.~Quarteroni, Accurate and efficient evaluation of failure
  probability for partial different equations with random input data, Computer
  Methods in Applied Mechanics and Engineering 267 (2013) 233 -- 260.

\bibitem{PEHERSTORFER201761}
B.~Peherstorfer, B.~Kramer, K.~Willcox, Combining multiple surrogate models to
  accelerate failure probability estimation with expensive high-fidelity
  models, Journal of Computational Physics 341 (2017) 61 -- 75.

\bibitem{doi:10.1137/17M1122992}
B.~Peherstorfer, B.~Kramer, K.~Willcox, Multifidelity preconditioning of the
  cross-entropy method for rare event simulation and failure probability
  estimation, SIAM/ASA Journal on Uncertainty Quantification 6~(2) (2018)
  737--761.

\bibitem{DIGE2018431}
N.~Dige, U.~Diwekar, Efficient sampling algorithm for large-scale optimization
  under uncertainty problems, Computers \& Chemical Engineering 115 (2018) 431
  -- 454.

\bibitem{LI2019106432}
M.~Li, Z.~Wang, Surrogate model uncertainty quantification for
  reliability-based design optimization, Reliability Engineering \& System
  Safety 192 (2019) 106432.

\bibitem{Ullmann2015MultilevelEO}
E.~Ullmann, I.~Papaioannou, Multilevel estimation of rare events, SIAM/ASA J.
  Uncertain. Quantification 3 (2015) 922--953.

\bibitem{ROCKAFELLAR2010499}
R.~Rockafellar, J.~Royset, On buffered failure probability in design and
  optimization of structures, Reliability Engineering \& System Safety 95~(5)
  (2010) 499 -- 510.

\bibitem{Sarykalin2008}
S.~Sarykalin, G.~Serraino, S.~Uryasev, {V}alue-at-{R}isk vs. {C}onditional
  {V}alue-at-{R}isk in Risk Management and Optimization, no. 2014, Institute
  for Operations Research and the Management Sciences (INFORMS), 2014, Ch.
  Chapter 13, pp. 270--294.

\bibitem{LEBRUN2009172}
R.~Lebrun, A.~Dutfoy, A generalization of the {Nataf} transformation to
  distributions with elliptical copula, Probabilistic Engineering Mechanics
  24~(2) (2009) 172--178.

\bibitem{LEBRUN2009312}
R.~Lebrun, A.~Dutfoy, An innovating analysis of the {Nataf} transformation from
  the copula viewpoint, Probabilistic Engineering Mechanics 24~(3) (2009)
  312--320.

\bibitem{ELDRED20111092}
M.~Eldred, L.~Swiler, G.~Tang, Mixed aleatory-epistemic uncertainty
  quantification with stochastic expansions and optimization-based interval
  estimation, Reliability Engineering \& System Safety 96~(9) (2011) 1092 --
  1113, quantification of Margins and Uncertainties.

\bibitem{Xie_DRCCP_Wassestein2018}
W.~Xie, On distributionally robust chance constrained programs with
  {W}asserstein distance, Mathematical Programming 186 (2021).

\bibitem{NANNAPANENI20169}
S.~Nannapaneni, S.~Mahadevan, Reliability analysis under epistemic uncertainty,
  Reliability Engineering \& System Safety 155 (2016) 9 -- 20.

\bibitem{Zaman17}
K.~Zaman, S.~Mahadevan, Reliability-based design optimization of
  multidisciplinary system under aleatory and epistemic uncertainty, Structural
  and Multidisciplinary Optimization 55 (2017).

\bibitem{Mourelatos06}
Z.~Mourelatos, J.~Zhou, A design optimization method using evidence theory, J
  Mechanical Design 128 (2006).

\bibitem{Youn08}
B.~Young, P.~Wang, Bayesian reliability-based design optimization using
  eigenvector dimension reduction {(EDR)} method, Structural and
  Multidisciplinary Optimization 36(2) (2008).

\bibitem{Hullermeier21}
E.~Hullermeier, W.~Waegeman, Aleatoric and epistemic uncertainty in machine
  learning: an introduction to concepts and methods, Machine learning 110
  (2021) 457--5006.

\bibitem{Campi21}
M.~Campi, A.~Care, S.~Garatti, The scenario approach: A tool at the service of
  data-driven decision making, Annual Reviews in Control 52 (2021) 1--17.

\bibitem{Paulson21}
J.~A. Paulson, A.~Mesbah, Data-driven scenario optimization for automated
  controller tuning with probabilistic performance guarantees, IEEE Control
  Systems Letters 5~(4) (2021) 1477--1482.

\bibitem{Campi21ml}
M.~Campi, S.~Garatti, A theory of the risk for optimization with relaxation and
  its application to support vector machines, Journal of Machine Learning
  22~(288) (2021).

\bibitem{Garatti22}
S.~Garatti, M.~Campi, Risk and complexity in scenario optimization,
  Mathematical Programming 191~(1) (2022) 243--279.

\bibitem{Crespo25}
L.~G. Crespo, B.~Stanford, N.~Alexandrov, A data-driven approach to risk-aware
  robust design, Reliability Engineering and System Safety.

\bibitem{Luedtke08}
J.~Luedtke, S.~Ahmed, A sample approximation approach for optimization with
  probabilistic constraints, {SIAM} Journal on Optimization 19~(2) (2008)
  674--699.

\bibitem{Nemirovski07}
A.~Nemirovski, A.~Shapiro, Convex approximations of chance constrained
  programs, {SIAM} Journal on Optimization 17~(4) (2007) 969--996.

\bibitem{Geng19}
X.~Geng, L.~Xie, Data-driven decision making in power systems with
  probabilistic guarantees: Theory and applications of chance-constrained
  optimization, Annual Reviews in Control 47 (2019) 341--363.

\bibitem{Mammarella22}
M.~Mammarella, V.~Mirasierra, M.~Lorenzen, T.~Alamo, F.~Dabbene,
  Chance-constrained sets approximation: A probabilistic scaling approach,
  Automatica 137 (2022) 110108.

\bibitem{Archimbaud18}
A.~Archimbaud, K.~Nordhausen, A.~Ruiz-Gazen, {ICS} for multivariate outlier
  detection with application to quality control, Computational Statistics and
  Data Analysis 128 (2018) 184--199.

\bibitem{Liang20}
Y.~Liang, A.~Thavaneswaran, Z.~Zhu, R.~K. Thulasiram, M.~E. Hoque, Data-driven
  adaptive regularized risk forecasting, in: 2020 IEEE 44th Annual Computers,
  Software, and Applications Conference, 2020, pp. 1296--1301.

\bibitem{Guan19}
L.~Guan, R.~Tibshirani, Prediction and outlier detection in classification
  problems, Journal of the Royal Statistical Society, Series {B} 84~(2) (2022).

\bibitem{Hanson10}
J.~Hanson, B.~Beard, Applying monte carlo simulation to launch vehicle design
  and requirement analysis, {NASA/TP} 2010 216447 (2010).

\bibitem{Campi25}
M.~C. Campi, A.~Care, L.~G. Crespo, S.~Garatti, F.~Ramponi, Risk analysis and
  robust design of data-driven models against adversarial actions, Journal of
  Machine Learning Research , submitted (2025).

\bibitem{Crespo08}
L.~G. Crespo, D.~P. Giesy, S.~P. Kenny, Robustness analysis and robust design
  of uncertain systems, AIAA Journal 46~(2) (2008).

\end{thebibliography}
 
\end{document}